\documentclass[twocolumn,amsmath,amssymb,prb]{revtex4-2}

\usepackage{color}
\usepackage{graphicx}
\usepackage{amsmath}
\usepackage{natbib}
\usepackage{dcolumn}
\usepackage{bm}
\usepackage{cases}
\usepackage{ulem}

\begin{document}
	\title{Finite-frequency conductivity of nonlinear Luttinger liquid in smooth random potential}

\author{A. A. Dontsov}
\email{operatorne@yandex.ru}
\affiliation{Ioffe  Institute, Politekhnicheskaya 26,  194021,   St.~Petersburg,   Russia}

\author{D. N. Aristov}
\affiliation{NRC ``Kurchatov Institute'' -- PNPI,   188300, Gatchina, Russia} 
\affiliation{Faculty of Physics, St. Petersburg State University, 199034, St. Petersburg, Russia}	
\affiliation{ Ioffe  Institute, Politekhnicheskaya 26,  194021,   St.~Petersburg,   Russia}

\author{A. P. Dmitriev}
\affiliation{ Ioffe  Institute, Politekhnicheskaya 26,  194021,   St.~Petersburg,   Russia}

\begin{abstract}
	We analyze the uniform conductivity of a one‑dimensional degenerate fermion system placed in a random disorder potential so smooth that backward scattering can be neglected. We use the nonlinear Luttinger liquid model to  consider effects of both interaction and the curvature of fermionic dispersion. The finite frequency conductivity, calculated in the lowest order of disorder potential, consists of two parts. First one is the elastic contribution, largely independent of temperature and interaction. Second one is the inelastic contribution, strongly dependent on temperature and frequency and appearing upon simultaneous presence of curvature, disorder and interaction.    We argue that apart from such finite frequency conductivity, there should always remain the $\delta$-function peak of conductivity at zero frequency, whose weight is weakly dependent on the disorder. 
%  Apart from that and few other minor constraints, the calculation is fairly general. For instance, we neither assume the interaction to be point-like nor weak. As a result, we obtain conductivity across a wide range of parameters, the parameters are the disorder scale, frequency, interaction radius and temperature.
\end{abstract}

\maketitle

\section{Introduction}

The advancements in nanotechnology allow one to create an increasing variety of types of 1D systems. Nowadays numerous types exist, including:  quantum wires, the chiral edge states of quantum Hall bars, the edge states of a two-dimensional topological insulator, carbon nanotubes etc. \cite{Deshpande2010}. A key aspect to note is that any point of these systems is accessible, allowing for direct measurement. The control of quasiparticles allows one to work with quantum information \cite{Bauerle2018}. Additionally, given the similarity between a photon in an optical waveguide and a quasiparticle in a 1D channel, the latter might parallel the role of quantum optical systems in basic research \cite{Bocquillon2014,Grenier2011}.

Luttinger liquid is one of the prevalent models of 1D systems in the limit of low energies \cite{Deshpande2010,Haldane1981}. However, many applications require taking into account the nonlinear dispersion \cite{Imambekov2012}. These applications include energy dissipation and relaxation \cite{apostolov2013thermal,Imambekov2012}, charge fractionalization \cite{Deshpande2010,das2011spin,dontsov2021charge}  and spin-charge fractionalization \cite{ma2017angle,Hashisaka2017,Giamarchi2003} , the Coulomb drag effect \cite{dmitriev2012coulomb,aristov2007,pustilnik2003coulomb}. Even the most subtle predictions of nonlinear Luttinger liquid theory are being confirmed experimentally \cite{glazman2020testing,moreno2016nonlinear}.

At zero temperature, nonlinear theory expands further, allowing the consideration of very high energy excitations. This expansion is the model of mobile impurity\cite{Imambekov2012,matveev2012scattering,pustilnik2006dynamic}. It was successfully used  in Hall systems \cite{mcginley2021elastic} in spin chain systems \cite{ovchinnikov2021threshold,schlottmann2019edge}, etc \cite{Imambekov2012} . Nonetheless, doubts exist regarding the model's ability to provide general order of accuracy. \cite{markhof2019investigating} 

The unusual nature of 1D systems is also demonstrated by their unusual quantum hydrodynamics at low temperature \cite{ruggiero2020quantum}. Many hydrodynamic properties have been calculated from first principles such as viscosity \cite{matveev2017viscous}. The problems of local thermalization relaxation \cite{buchhold2016prethermalization,ristivojevic2013relaxation,matveev2013equilibration} and the thermal front propagation \cite{bertini2018universal} are also actively studied.  The model of mobile impurity also allows performing such calculations \cite{matveev2012scattering}.

This work is concentrated on fermions, however, it is important to note that 1D boson systems are also actively studied experimentally \cite{schlottmann2018exponents,bertaina2016one}, and theoretically \cite{lang2018correlations,fabbri2015dynamical}

Transport properties of a Luttinger liquid have been the focus of considerable research due to their unusual nature compared to three dimensional systems. Owing to one-dimensionality, a small impurity radically changes the ideal ballistic character of the particles motion \cite{kane1992}, and even a weak local cluster of impurities dramatically affects the general conductance as well as the tunneling conductance \cite{polyakov2003transport,das2019conductance,das2019transport}. On the other hand, the fact that a relatively short system has electrical leads becomes crucial in one dimension  \cite{maslov1995dirtyandleads}. 

The case opposite to a single point-like impurity is the case of smooth disorder\cite{kopietz1999optical,maslov1995dirtyandleads}.  It was shown that nonlinearity of the spectrum and forward scattering on the random potential are crucial in this case \cite{kopietz1999optical}. In  case of zero-frequency limit, the  conductance and other transport properties were considered using the kinetic equation method \cite{levchenko2010transport} as well as hydrodynamics \cite{andreev2011hydrodynamic}. Additionally, this case of smooth disorder is interesting because the effect of weak localization can take place \cite{gornyi2007electron, gornyi2005interacting}.

This paper considers the conductivity $\sigma(\omega)$ of a one-dimensional degenerate liquid system of electrons with a finite mass $m$ in a sample with smooth disorder $U(x)$, whose scale, $d$, is assumed large and the potential is assumed weak, $U(x)\ll E_F$, where $E_F$ is the Fermi energy. 
% In the present paper the bosonization technique is used.  For our problem, this technique permits one to consider only frequencies bigger than some minimal limit, but on the other hand, it allows one to find the conductivity from the first principles with minimal other restrictions, for instance, the interaction between the particles does not have to be point-like nor weak, see Appendix \ref{sec:AREAQuantum}. A forthcoming article will use the kinetic equation method. It does not limit the frequency from below, only from above and the interaction has to be weak. Within the area where both methods are applicable, the conductivity obtained with these methods coincide exactly. 

We employ bosonization technique and consider  nonlinear fermionic dispersion, which corresponds to decay of plasmons in bosonization. 
% A system with nonlinear dispersion is considered. 
The system is assumed long enough for the effect of the leads to be omitted. As usual for Luttinger liquid, we let $a\gg\lambda_F$, where $a$ is the range of interaction and $\lambda_{F}$ is the Fermi wavelength. Also, we assume that the random potential is large-scaled, meaning $d\gg a$ and, thereby, $d\gg \lambda_F$; this allows us to neglect backward scattering from the random potential. Apart from these spatial values, there are two more: the thermal length, {\color{black} $l_T=v_F/ T$}, and the frequency length, $l_\omega=\tilde{v}/\omega$, where $T$ is the temperature, 
% $k_{F}$ is the Fermi momentum 
and $\tilde{v}$ is plasmon velocity
% a renormalized Luttinger velocity different from $v_{F}$, density fluctuations move at this velocity. 
Different relations between these lengths correspond to different regimes (frequency and temperature dependence) of conduction. Throughout the paper, we assume that the temperature is low enough $\lambda_F\ll l_T$, hence the liquid is degenerate. Summing up, the conditions for length scales are
\begin{eqnarray}
	\label{eq:lengthCondits}
	d\gg a\gg \lambda_F;\quad l_T\gg \lambda_F.
\end{eqnarray}	
Note that  non-uniformity is necessary for the dissipation to be nonzero $\sigma(\omega)\neq0$, when $\omega\neq0$.

% It can be explained by the fact that the response of a system to a uniform force field is purely inertial and independent of inner interactions between the particles. Indeed it is nothing but a center-of-mass motion caused by an applied force uniform in space \cite{Imambekov2012}. 

The rest of the paper is organized as follows.	
We set up our problem and derive a general formula which expresses the conductivity through the electron liquid density and the random potential in Sec.\ \ref{sec:setup}.
Considering a gas of non-interacting electrons, we obtain expressions for the conductivity in the classical consideration and in the bosonization treatment in Sec.\ \ref{sec:noninter}, establishing coincidence of the results in two approaches.  
The density correlation function without random potential is described in Sec.\ \ref{sec:Correlat} for interacting system.  
The conductivity of interacting 1D liquid is analyzed in Sec.\ \ref{sec:inter}, where we show the existence of elastic and inelastic contributions to optical conductivity. We briefly summarize the main results of the work in the concluding section \ref{sec:Conclusions}.
Finally, in Appendix \ref{sec:NonIntClassi} we provide the classical derivation of the conductivity of  fermion gas in smooth disorder potential.

% Setup of the problem} 
% conductivity of  non-interacting Fermi gas
% density correlations in interacting case
% The conductivity of  interacting 1D liquid
% One-particle contribution obtained with the classical Liouville's equation

% In Section \ref{sec:inter}, we consider the gas of interacting electrons within the framework of the bosonization method and obtain specific expressions for the conductivity at  different ratios between the spatial scales.

\section{\label{sec:setup} Setup of the problem} 
We consider  the 1D Hamiltonian  
\begin{equation}
	\begin{aligned}
		{\cal H} &=%	 {\cal H}_0 + \cal{H_{U}} + {\cal H}_{int} \,, \quad {\cal H}_0 = 
		\int dx\,   ( H_0  +H_{U} +H_{int}  ) \,,    \\  
		H_0  & =  
		\pi v_{F} ( R^{2}(x)+L^{2}(x) )  + \tfrac{2 \pi^{2}}{3 m} (   R^{3}(x)+L^{3}(x) ) \,, \\ 
		H_{U} &=   U(x) \rho(x) \,,  \\
		{H}_{int}  & =   \frac{1}{2} \int  dy\,  {\color{black} \delta}\rho(x)\, g(x-y) \,{\color{black} \delta}\rho(y)    \,,
	\end{aligned}
	\label{eq:HamilDens}
\end{equation}
where $g(x-y)$ is the interaction between particles. Here we introduced smooth right- and left-moving  particle densities,  $R(x)$   and $L(x)$, and the smooth part of total density,  $\rho(x)=R(x)+L(x)$. {\color{black} Requiring electroneutrality, we write $\delta\rho(x)=\rho(x) - \langle\rho(x)\rangle $, with the average density $\langle\rho(x)\rangle \simeq -U(x) /\pi v_{F}$ is found by classically minimizing ${\cal H}$ with respect to $R$ and $L$, see also Appendix \ref{AppOPE}.  }
% {\color{black}, along the lines of the bosonization approximation}. 
We assume that the smooth random potential, $U(x)$,  is characterized by the only correlator \[ W(x_2-x_1)= \overline{U(x_2) U(x_1)} \,, \] while $ \overline{ U(x)} = 0$, here overbar stands for averaging in disorder realizations.

The Tomonaga commutation relations between the operators are 
\[\left[ {R}(x),  {R}\left(x^{\prime}\right)\right]=-\left[ {L}(x),  {L}\left(x^{\prime}\right)\right]=\tfrac{i}{2 \pi} \partial_{x} \delta\left(x-x^{\prime}\right) \, , \] and $ \left[ {R}(x),  {L}\left(x^{\prime}\right)\right]=0$. 

The cubic terms in fermionic densities in \eqref{eq:HamilDens}, $\propto 1/m$, appear in bosonization description  thanks to  quadratic curvature of the fermionic spectrum, see \cite{aristov2007} and references therein.

% \section{\label{sec:condThrougCorr} A general formula for the conductivity}

% In this section using the bosonization method we obtain a general formula  that expresses the conductivity through the electron liquid density and the random potential. In the case of a weak potential using the formula we derive a simpler one that defines the conductivity through the random potential correlator $W(x_2-x_1)=\left\langle U(x_2) U(x_1) \right\rangle $ and through the electron liquid density correlator of a liquid without the external potential $\mbox{Im}\left\langle\rho \rho\right\rangle_{q, \omega}$.

The time evolution,  $\partial_{t} R  =i[{\cal H}, R]$,   is given by 
\begin{equation}
	\label{eq:RLopersDiff}
	\begin{aligned} 
		\partial_{t} R(x, t) &  =-\partial_{x}[ v_{F} R(x, t)+\frac{\pi}{m} R^{2}(x, t) \\
		&+\frac{1}{2 \pi} \int d y\,  g(x-y) \,{\color{black} \delta}\rho(y, t) +\frac{1}{2\pi} U(x) ]  \,, \\ 
		\partial_{t} L(x, t) &  =\partial_{x}[ v_{F} L(x, t)+\frac{\pi}{m} L^{2}(x, t) \\
		&+\frac{1}{2 \pi} \int d y\,  g(x-y) \,{\color{black} \delta}\rho(y, t) +\frac{1}{2 \pi} U(x)] \,.
	\end{aligned} 
\end{equation}		 
From the  continuity equation, $\partial_{t} \rho+\partial_{x} j=0$, we find  the current operator 
\begin{equation}
	\label{eq:CurrentDiff}
	j=v_{F}(R-L)+\frac{\pi}{m}\left(R^{2}-L^{2}\right).
\end{equation}

We are interested in calculation of the electric conductivity, which can be 
represented in Kubo formalism as the response function of the current, $j(x, t)$. We write it for $\omega\neq0$ in {\color{black} the form \cite{Rosch2005,aristov2007} }
\begin{equation}
	\label{eq:Kubo}
	\sigma(\omega)= -\frac{e^2\,\mbox{Im}\langle\!\langle \overline{\partial_{t} j ,\, \partial_{t} j} \rangle\!\rangle |_{q=0}}{\hbar\, \omega^{3}} \,.
\end{equation}
Here and below we denote  the retarded Green's function
in $(q,\omega)$-representation  by the double angular brackets,  
\[
\langle\!\langle  A,\, B \rangle\!\rangle =  -i  \int_{0}^{\infty} dt\int dx\, \left\langle[A(x,t),\,B(0,0)]\right\rangle  e^{- i  q x+ i  \omega t} \,,
\]
where angular brackets stand for quantum-thermodynamic averaging.  

% To use the equation of continuity
%\begin{equation}\label{eq:Nepr}	\partial_{t} \rho+\partial_{x} j=0 \end{equation}
% for obtaining the particle flux, find a derivative of the density operators:

% We note that those terms in $\partial_{t} j(x,t)$ , which are full derivatives on $x$ , do not contribute to (\ref{eq:Kubo}) in the uniform limit, $q=0$.

Calculating now $\partial_{t} j(x, t)=i[H, j(x, t)]$, and using Eq. (\ref{eq:RLopersDiff}),  we notice that the linear-in-densities terms in (\ref{eq:CurrentDiff}) give a full derivative with respect to $x$ and, thus, can be omitted in calculation of $\sigma(\omega)$ in the uniform limit, $q=0$. These terms correspond to usual Tomonaga-Luttinger theory. Similarly, commuting the current with    $H_{0}$ in \eqref{eq:HamilDens} produces the full derivative by $x$ and can also be omitted.  We obtain eventually
% \begin{equation}  \begin{aligned}  	
		\[ \partial_{t} j(x, t)  =-\frac{\rho(x, t)}{m}\left[ \partial_{x} U(x)+ \int \partial_{x} g(x-y) \,{\color{black} \delta}\rho(y, t) d y\right] \,. \]
		% \end{aligned} \label{eq:dtj} \end{equation}		
%This expression means that the time derivative of the current is given by the sum of two forces: the first one is the density multiplied by the local force created by the disorder; the second one is the force created by the other electrons. 

The Fourier transform of the second term above {\color{black} partly} vanishes in the limit of our interest, $q=0$,   since $\iint d y d x\, \partial_{x} g(x-y) \rho(y, t) \rho(x, t)=0$. Indeed, the derivative of the interaction potential is antisymmetric, $\partial_{x} g(x-y)=-\partial_{x} g(y-x)$, and  the combination $\rho(y, t) \rho(x, t)$, taken at the same time, is symmetric in $x,y$. 

 {\color{black}  The non-vanishing part of the second term is 
 $-\int \partial_{x} g(x-y) \, \langle\rho(y,t)\rangle d y \simeq \partial_{x}\int dy\, g(x-y) \,  U(y)/ (\pi v_{F})$. In view of the property  $d\gg a$, we may represent this contribution as $  \partial_{x}    U(x) / (\pi v_{F}) \int  dy\, g(y)$. Using the notation for the Luttinger parameter, $K =1/\sqrt{1+g_0/(\pi v_F)}$, introduced below, we conclude that  
}
%It means that 
for our purposes, it suffices to consider a following expression
\begin{equation} \partial_{t} j(x, t)=-\tfrac{1}{m {\color{black}  K^{2} } } \rho(x, t) \partial_{x} U(x) \,.  
\label{dJdt}
\end{equation}
The above  formula for the conductivity takes the form  
\begin{equation}
	\begin{aligned}
    \label{eq:generCond}
		\sigma(\omega) & =-\frac{e^2 \, {\color{black}  K^{-4} } }{\hbar\, m^{2} \omega^{3}} \mbox{Im}\langle\!\langle  
		\overline{(\partial_{x} U) \rho ,(\partial_{x} U)\rho } \rangle\!\rangle _{q=0} \,.
	\end{aligned}
\end{equation}

{\color{black} In the absence of random potential, there is no conductivity at $\omega \neq 0$. A nonuniform random potential breaks this property and leads to a nonzero real part of the conductivity.  Indeed, the relation \eqref{eq:generCond} shows that only the potential gradient $\partial_x U$ is relevant here. 
 It also means a constant potential does not produce finite-$\omega$ conductivity and the conductivity is gauge-consistent, as it should be.  See also Appendix  \ref{AppOPE}. }

When deriving formula \eqref{eq:generCond},  we have not made any additional assumptions. Assuming now that the magnitude of the random potential $U(x)$ is small,  we   obtain the  conductivity  in the lowest order 
\begin{equation}
	\begin{aligned}
		\label{eq:DmitSigmaGENERAL}
		\sigma(\omega) & = \frac{e^2 {\color{black}  K^{-4} } }{ \hbar\,m^{2} \, \omega^{3}} 
		\int  \frac{d q}{2 \pi} q^{2}  {W}_q   {\cal D}_{q,\omega}  \,, \\ 
		% \sigma(\omega) & =-\frac{e^2}{ \hbar\,m^{2} \, \omega^{3}} \int  \frac{d q}{2 \pi} q^{2}  {W}_q \mbox{Im}\langle\!\langle\rho , \rho\rangle\!\rangle  \,, \\ 
		{\cal D}_{q,\omega}  & =  - \mbox{Im}\langle\!\langle\rho , \rho\rangle\!\rangle_{q,\omega} \,, \\ 
		{ W}_{q} & = \int d x\, e^{-iqx} W(x) \,. 
	\end{aligned}
\end{equation}
% where %  the angular brackets denote thermodynamic averaging in the clean system ($U\equiv0$) and 
% $  { W}_{q} = \int d x\, e^{-iqx} W(x)$.  
We emphasize here that expression \eqref{eq:DmitSigmaGENERAL} stems from the second term in   $H_{0}$, Eq. \eqref{eq:HamilDens}, which is responsible for the nonlinearity of the fermion spectrum. In case of linear fermionic dispersion, $1/m=0$, we have $\sigma(\omega)=0$ at any $\omega\neq0$. 

%{\color{black} (moved here from the end of the paragraph)}
The calculation of conductivity is thus reduced in the leading order to evaluation of the spectral density $ {\cal D}_{q,\omega} $ for the interacting electron gas without  random potential. The structure of $ {\cal D}_{q,\omega} $ for a point-like interaction is known, especially for zero temperature \cite{aristov2007,Imambekov2012}.  In this paper we obtain $ {\cal D}_{q,\omega} $  for finite-range interaction, $g(x)$ 
The range of $q$ contributing to the conductivity is limited by the smoothness of potential, as 
the value of $ {W}_q$ is expected to decrease rapidly (faster than any power) with $q$ beyond $1/d$. It restricts  integration in \eqref{eq:DmitSigmaGENERAL} by $q\lesssim 1/d \ll k_F$.

In  paper \cite{aristov2007} the interaction between two one-dimensional particle systems is considered. If one let the mass of particles in one of the wires infinite and their velocity tending to zero, then they freeze and induce a random large-scaled potential on the second wire. In this limit the previous and the present problem become similar.

\section{\label{sec:noninter} %The  density correlator and 
	conductivity of  non-interacting Fermi gas }

\subsection{\label{subsec:noninterClassic} The classical conductivity  of  non-interacting gas }

It is instructive to perform first a classical consideration of non-interacting Fermi gas.

We write the classical Liouville's equation for the right-moving electrons 
\begin{equation}
	\frac{\partial f_{R}}{\partial t}+\frac{\partial f_{R}}{\partial x} \frac{\partial H}{\partial p}-\frac{\partial f_{R}}{\partial p} \frac{\partial H}{\partial x}=0,
\end{equation}
with $H=\frac{p^2}{2 m}+U(x)+e {\cal E}_{0} x \cos(\omega t)$ and ${\cal E}_{0}$ is  the electric field. We rewrite this equation in terms of a new variable 
$E=p^2/2m+U(x)$, in the following way
\begin{equation}
	\begin{aligned}
		\frac{\partial f_{R}}{\partial t} & +\sqrt{\frac{2}{m}(E-U(x))} \frac{\partial f_{R}}{\partial x}
		\\ &=e {\cal E}_{0} \exp (i \omega t) \sqrt{\frac{2}{m}(E-U(x))} \frac{\partial f_{R}}{\partial E}   \,.
	\end{aligned}
	\label{eq:KinurClass}
\end{equation}
Assuming that the  field ${\cal E}_{0}$ is weak, we replace $f_{R}$ in the right hand side of \eqref{eq:KinurClass} by the Fermi distribution function of the right electrons, $f^{(0)}_{R}$, and $f_{R}$ in the left hand side now means deviation from $f^{(0)}_{R}$. The equation for the left electrons is similar, with $f_{ R}$ being replaced by $f_{ L}$, and a different sign before $\sqrt{\frac{2}{m}(E-U(x))}$. Eq.\ \eqref{eq:KinurClass} is solved  for $T=0$ in Appendix \ref{sec:NonIntClassi}. The real part of the obtained classical conductivity  is
% {\color{black} 
% \sout{\begin{equation}\label{eq:ClassResult}
% 	\begin{aligned}
% 		\sigma_{\text{cl}}(\omega)& \simeq \frac{e^{2} v_{F}}{\pi \hbar}\left (1-\frac{3 W(x=0)}{2 m^{2} v_{F}^{4}}\right )  \delta(\omega)
% 		 +\frac{e^{2} W_{ q= \omega / v_{F} }}{2 \pi \hbar \, m^{2} v_{F}^{4}}.
% 	\end{aligned}
% \end{equation}}

\begin{equation}\label{eq:ClassResult}
	\begin{aligned}
		\sigma_{\text{cl}}(\omega)& \simeq \frac{e^{2} v_{F}}{ \hbar}\left (1-\frac{3 W(x=0)}{2 m^{2} v_{F}^{4}}\right )  \delta(\omega)
		\\ & +\frac{e^{2} W_{ q= \omega / v_{F} }}{2 \pi \hbar \, m^{2} v_{F}^{4}}. 
	\end{aligned} 
\end{equation}

The physical reason for nonzero conductivity at $\omega\neq0$ in a noninteracting liquid is the following. In the absence of disorder, under the influence of an oscillating uniform electric field, the electron velocity oscillates with the same frequency and with a  phase shift of  $\frac{\pi}{2}$. As a result, the work of the field during the period becomes zero, which means $\sigma(\omega\neq0 )=0$. Under the influence of disorder, the phase shift between the field and the velocity becomes irregular and the conductivity turns out to be finite.

We note that from physical considerations, it is clear that in our setup and in a constant electric field ($\omega = 0$), the speed of electron  increases infinitely and the conductivity becomes infinite.  Note that  the conductivity of a one-dimensional system of non-interacting electrons without backward scattering was studied in  \cite{kopietz1999optical}, where the $\delta$-function part of conductivity was simply omitted. The authors drew an unsubstantiated conclusion that the smooth disorder in 1D systems completely smears the $\delta$-function in the framework of the bosonization approach. %We show below that the $\delta$-function is smeared but because of a completely different reason.  

%$\mbox{Im}\left\langle\rho \rho\right\rangle_{q,\omega}$

\subsection{\label{subsec:noninterQuantum} The quantum conductivity of   non-interacting gas }

% \subsection{ The case without interaction}
\label{subsec:correlNonInter}

\begin{figure}
	\centering
	\includegraphics[width=0.7\linewidth]{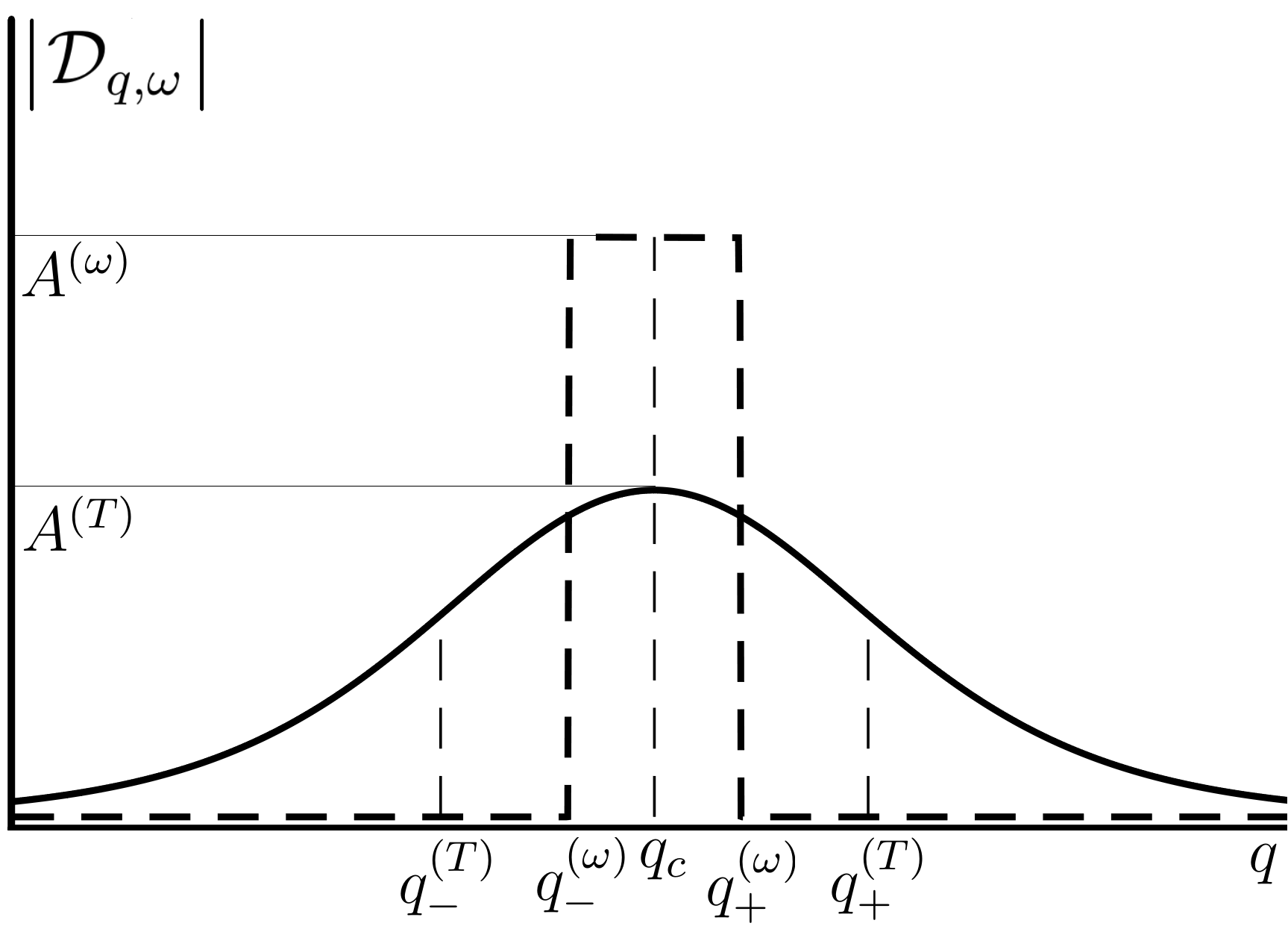}
	\caption{Correlator ${\cal D}_{q,\omega}$ for a non-interacting gas with nonlinear dispersion as a function of $q$ with fixed $\omega$, schematically shown for low temperature $T\ll \omega$  (dashed line) and for relatively high  temperature $T\gg \omega$ (solid line). The low temperature peak is located between the values $q^{(w)}_-$ and $q^{(w)}_+$  and the high temperature peak is mostly present between $q^{(T)}_-$ and $q^{(T)}_+$. The heights of the peaks are denoted by $A^{(w)}$ and $A^{(T)}$, respectively. }
	\label{fig:RRnonInt}
\end{figure}

In the limit of linear dispersion, $1/m\rightarrow0$, and without interaction, one has a well-known expression  \cite{Giamarchi2003} 
% see Eq. (2.61) there 
% \begin{equation}\label{eq:rrNoInt} 
	% 	\mbox{Im} \langle\!\langle\rho, \rho \rangle\!\rangle = -\tfrac12 { q} \left[\delta\left(\omega - q v_{F} \right)-\delta\left (\omega + qv_{F}\right)\right].
	% \end{equation}
\begin{equation}\label{eq:rrNoInt} 
	{\cal D}_{q,\omega} = \tfrac12 { q} \left[\delta\left(\omega - q v_{F} \right)-\delta\left (\omega + qv_{F}\right)\right] \,.
\end{equation}
In  case of nonlinear dispersion and zero temperature the  density correlator is finite at all $q$ and has \cite{Imambekov2012} a rectangular shape  \footnote{ The dynamical structure factor $S_q(\omega)$ usually studied is proportional to ${\cal D}_{q,\omega}$ by means of fluctuation-dissipation theorem and one can talk of either one. }; for  finite temperatures it is a broadened temperature function \cite{aristov2007}
\begin{equation}\label{eq:rrNoIntNonLinear}
	\begin{aligned}
		{\cal D}_{q,\omega} 
		% \mbox{Im}\langle\!\langle\rho, \rho \rangle\!\rangle 
		% & =-\frac{m}{4 q}\frac{ \sinh \left(\frac{\omega}{2 T}\right)}{ \cosh \left(\frac{ v_F \left[2 m \left\lbrace \omega-q v_F\right\rbrace-q^2\right]}{4 T q}\right) \cosh \left(\frac{ v_F \left[2 m \left\lbrace \omega-q v_F\right\rbrace+q^2\right]}{4 T q}\right) }, \\
		& = \frac{m}{4 q}\frac{ \sinh ({\omega}/{2 T} ) }
		{ \cosh  \frac{     \omega-q v_F -q^2/2 m }{2 T q/ (m v_F)}  
			\cosh  \frac{      \omega-q v_F +q^2/2 m }{2 T q/ (m v_F)}  }, 
	\end{aligned}
\end{equation}
which is shown in Fig. \ref{fig:RRnonInt}. The center of this spectral weight function is at $q_c=\omega/v_F$. 
% {\color{black} ( The figure shows positive $q_{c}$ and hence positive $\omega$, the above formula corresponds to negative amplitude of the peak. The imaginary part of retarded function equals to spectral weight with minus sign.) } ({\color{black}  replaced the correlator with its modulus in the figure})
Its width is  estimated as $|q_\pm^{(\omega)}-q_c|\approx  q_c^2/2m v_F$ at low temperatures, $T\ll \omega$ and at relatively large  temperature, $T\gg \omega$, the width is $|q_\pm^{(T)}-q_c|\approx \frac{q_c^2}{2 m v_F} \frac{T}{\omega}$. Overall we may write 
\begin{equation}
	\label{eq:qplusqminus}
	\begin{aligned}
		q_+-q_-\approx \frac{q_c^2}{ m v_F} \frac{\max (\omega,T)}{\omega}.
	\end{aligned}
\end{equation}
The height of the peak is given by $A\approx\frac{v_F\, m}{\max (\omega,T) }$.

%%%%%%%%%%%%%%%%%%%%%%%
%%%%%%%%%%%%%%%%%%%%%%%

The calculation of the  conductivity of a non-interacting 1D gas with nonlinear dispersion may be done with the full nonlinear correlator (\ref{eq:rrNoIntNonLinear}) in Eq.\  (\ref{eq:DmitSigmaGENERAL}). 
However,  we can safely
% As it was stated in Sec. \ref{subsec:correlNonInter} the possibility to 
replace the complicated shape of the  peak (\ref{eq:rrNoIntNonLinear}) concentrated between $q_+$ and $q_-$ by a simple delta function (\ref{eq:rrNoInt}), given the large-scale character  of the external potential, $1/d\gg (q_+-q_-)$. The latter inequality is easily satisfied in the following frequency domain
%	 \begin{eqnarray}
	%	 	\label{eq:LargeScale}
	%	 	\omega\ll \frac{v_F}{d} \frac{m v_F^2}{\max(T,\hbar \omega)}.
	%	 \end{eqnarray}	
%	 \begin{eqnarray}
	% 	\label{eq:LargeScale}
	%	\omega\ll \frac{v_F}{d} \frac{m v_F^2}{\max(T,\hbar \omega)}.
	% \end{eqnarray}	
\begin{equation}
\label{eq:LargeScale}
\begin{cases}
%	\omega \ll \frac{v_F}{d} \frac{m v_F^2}{T} \,,  \quad    \hbar\omega \ll T \,,\\
%	\omega \ll \sqrt{\frac{v_F}{d} \frac{m v_F^2}{\hbar}} \,,  \quad  \hbar \omega \gg T \,.
	\omega \ll   E_{F} \,\omega_d/ T\,,  \quad    \hbar\omega \ll T \,,\\
	\omega \ll \sqrt{ E_{F}\, \omega_d} \,,  \quad  \hbar \omega \gg T \,,
\end{cases}    
\end{equation}
with $\omega_d = \hbar \tilde{v} /d$ and  $E_{F} = m v_F^2/2$. 

Using Eq.\  \eqref{eq:rrNoInt} in \eqref{eq:DmitSigmaGENERAL}  we obtain 
\begin{equation}\label{eq:NonInteracResult}
	\sigma_{1}(\omega)=\frac{e^{2} W_{ q= \omega / v_{F} }}{2 \pi \hbar\,  m^{2} v_{F}^{4}}.
\end{equation}
This expression coincides  with the classical result, Eq. (\ref{eq:ClassResult}),  at $\omega \neq 0$. The explanation is simple: the quantum parameter $\lambda_F$ is   smaller than the scale of the disorder, $d\gg \lambda_F$, which means the applicability of the semiclassical approach. 

{\color{black}  We note that the one-boson result does not contain an explicit $1/\omega^2$ frequency dependence. 
However, the one-boson contribution still decreases with frequency, since $W_\omega$ decreases exponentially when $\omega/v \gg 1/d$. 
The factor $1/\omega^2$ is canceled by the small factor $q^2$ in the numerator, because the delta function fixes $q=\omega/v$. In its turn, the factor $q^2$ in \eqref{eq:DmitSigmaGENERAL} in our 1D case is similar to the transport life-time factor, $ \propto (1-\cos \theta) \sim \theta^2$, in higher spatial dimensions, which filters out forward scattering processes.
}
% The  applicability of this calculation for quantum case is discussed in Appendix \ref{sec:AREAQuantum}.

\section{\label{sec:Correlat} density correlations in interacting case
	%	$\mbox{Im}\left\langle\rho \rho\right\rangle_{q, \omega}$
}

% To find the conductivity, as indicated by (\ref{eq:DmitSigmaGENERAL}), it is enough to calculate the imaginary part of the retarded Green's function $Im\left\langle\rho \rho\right\rangle_{q,\omega}$. In the present section we list several formulas for it. 

\subsection{ General remarks }
% \subsection{ Review of the interacting correlator structure}
\label{subsec:correlINTERACTINGpointlike}

Let us  describe the changes in  density correlation function in  presence of the interaction and  in  case of nonlinear fermionic dispersion. First,  the non-perturbed Fermi velocity, $v_{F}$, is replaced by plasmon velocity, $v$, in the corresponding formulas. Second, the simple peak in Fig. \ref{fig:RRnonInt} warps in a complex way \cite{Imambekov2012}, it is now concentrated near $q_c=\omega/v$ between $q_\pm=q_c\pm \frac{q_c^2}{2 m v} \frac{\max (\omega,T)}{\omega}$, its height is  $A\approx\frac{v_F\, m}{2 \,\max (\omega,T) }$, it  weakly depends on the interaction. Third, besides the changes in the peak, long tails are formed as shown in Fig. \ref{fig:ImRR}.  These tails decrease slowly and, as shown below, provide a major contribution into the overall conductivity (\ref{eq:DmitSigmaGENERAL}) for some sets of parameters.  

%%%%%%%%%%%%%%%%%%%%%
\begin{figure}
	\centering
	\includegraphics[width=0.7\linewidth]{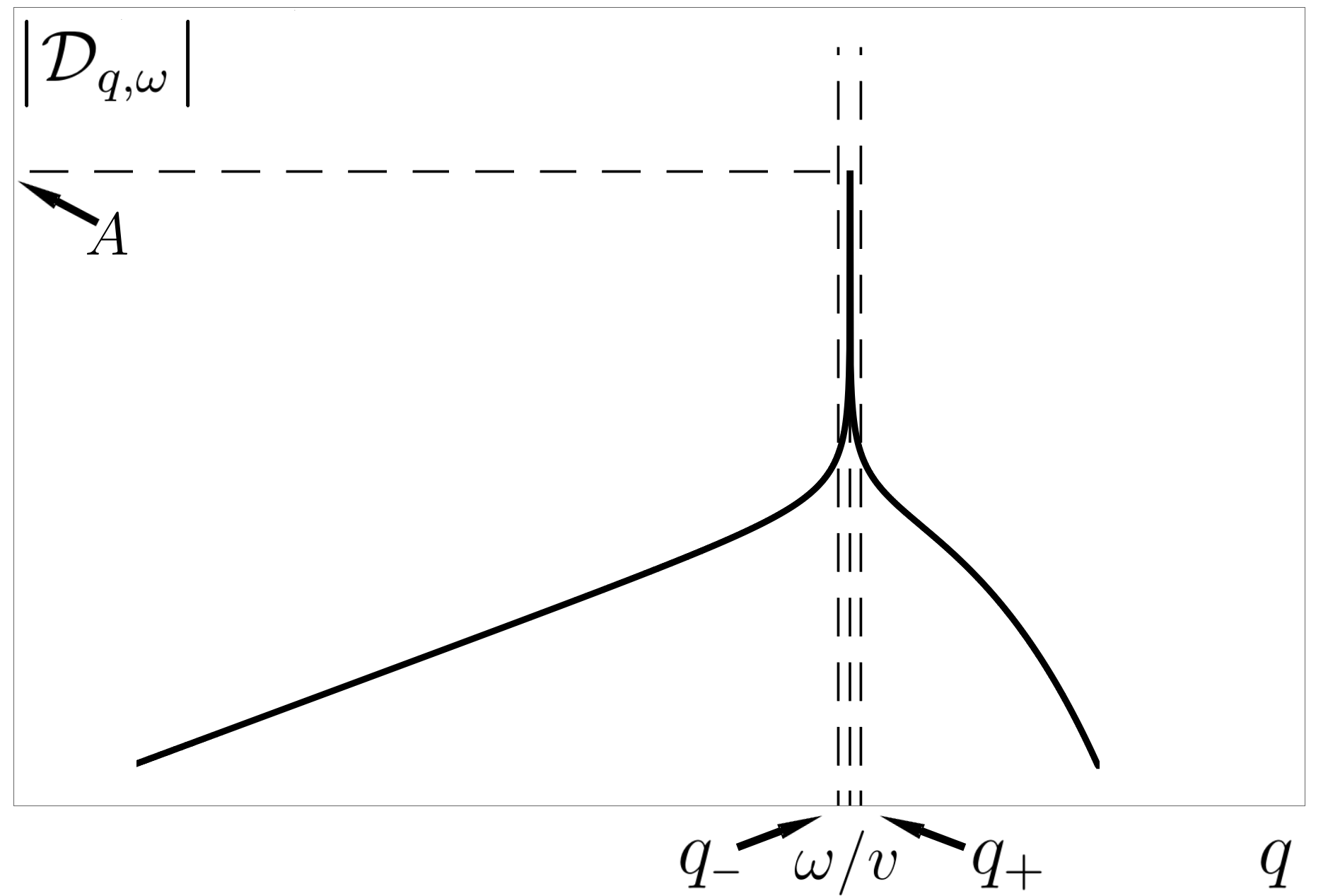}
	\caption{Correlator   $ {\cal D}_{q,\omega} $ as a function of $q$ with a fixed $\omega$ in log-log format. The temperature is chosen low $T\ll \tilde{v}/d$. The warped  peak from Fig. \ref{fig:RRnonInt} is concentrated between $q_\pm$, see   Eq. \eqref{eq:qplusqminus}. Interaction induces long tails, that may decrease very slowly. This picture is drawn for a point-like interaction, for a finite range interaction the picture is similar.  
		\label{fig:ImRR} }
\end{figure}
%%%%%%%%%%%%%%%%%%%%%

% \subsection{\label{subsec:correlINTERACTING_NON_pointlike} The calculation of a finite-range interaction correlator.}

% As noted at the end of Sec. \ref{sec:condThrougCorr}, the applicability condition for the Tomonaga-Luttinger model, $q\ll k_F$, is always satisfied. Therefore, it seems reasonable to 
We consider the nonlinearity of the fermionic dispersion as a small perturbation of  Tomonaga-Luttinger model at $q\ll k_F$. The cubic in density terms of the Hamiltonian, $\propto 1/m$, are strongly irrelevant perturbation, that is they cannot modify the spectrum, although they can modify observables. 
Within such perturbation theory, the results of interacting Tomonaga-Luttinger model are obtained as the leading term, ${\cal O}(m^{0})$, in the asymptotic series in powers of $1/m$. 
% The nonlinear term, $\propto 1/m$, in the Hamiltonian is a small perturbation.
% Generally, this perturbative approach fails and the diagrams for $\langle\!\langle\rho, \rho\rangle \!\rangle$ diverge  \cite{Imambekov2012}. 
This asymptotic series of contributions to $\langle\!\langle\rho, \rho\rangle \!\rangle$ is not converging (even in the  free-fermion case), as there is no way to obtain the spectral weight of finite width, Eq. \eqref{eq:rrNoIntNonLinear}, starting from $\delta$-functions, ${\cal D}_{q,\omega} $ in Eq.\ \eqref{eq:rrNoInt}. 
The finite-width spectral weight is obtained by associating a part of  bosonized Hamiltonian with composite fermions having non-linear dispersion  \cite{Rozhkov2005,Imambekov2012}. 
%,  or use certain regularization \cite{aristov2007}. 
%The mathematical reason here is that the linear approximation for fermionic dispersion is only able to give a delta function for $\mbox{Im}\langle\!\langle\rho, \rho\rangle \!\rangle$ as a function of $q$, but the real correlator is a wrapped rectangle or a wide curve (\ref{eq:rrNoIntNonLinear}) of the width $\sim1/m$, and no series is able to transform one into another \cite{Imambekov2012}. 

We find, however,  that corrections to the imaginary part of Green's function ${\cal D}_{q,\omega} $ away from the peak position, $q\simeq \omega/v$, 
% \footnote{This point is actually the pole of the first correction, therefore, while near the pole one has to strictly resum the diagrams, far from the pole all diagrams except the first correction are irrelevant.} 
are important in our calculation of conductivity. At the same time, the  details of the peak's profile can well be absorbed into its cumulative characteristics, such as the position and the magnitude. 

%this perturbation theory is applicable and no regularization is needed \cite{pereira2007dynamical,aristov2007}. This is the approach we adopt in this work.

%Exploring these ideas, we see that the correlator naturally splits into two parts, 
For our purposes, it is thus natural to divide the density spectral weight into two parts,  
the Tomonaga-Luttinger correlator as a zero approximation and corrections to it.  We write  ${\cal D}_{q,\omega}  \simeq {\cal D}_{q,\omega} ^{(0)}+{\cal D}_{q,\omega} ^{{(1)}}$, 
where ${\cal D}_{q,\omega} ^{{(0)}}$ does not depend on $m$ and 
${\cal D}_{q,\omega} ^{{(1)}}\sim  m^{-2}$. 
The Tomonaga-Luttinger part,  ${\cal D}_{q,\omega} ^{{(0)}}$,  approximates the peak at  $q_c \simeq\omega/v$  by $\delta$-function, which is sufficient for our calculation { (see Fig. \ref{fig:RRnonInt})}. The correction  ${\cal D}_{q,\omega} ^{{(1)}}$ describes the  tails stretching { far} away from the point $q_c=\omega/v$, we show  this correlator  in Fig. \ref{fig:ImRR}.

We refer to the corresponding contributions in the correlator,   ${\cal D}_{q,\omega} ^{ {(0)}}$ and 
${\cal D}_{q,\omega} ^{ {(1)}}$,  as single-boson  and  double-boson contributions, respectively. 
This terminology is due to the fact, after the diagonalization of the Tomonaga-Luttinger Hamiltonian, the new chiral bosons do not interact. The combination of fermionic interaction and non-linearity of  the fermionic spectrum leads to possibility of new chiral bosons decaying into two bosons of different chirality in the intermediate state. 

%The interaction between the bosons is proportional to $1/m$.

\subsection{One- and two-boson contribution to density  correlator}

In the linear dispersion limit  a standard calculation  \cite{Giamarchi2003}  gives 
$\langle\!\langle\rho, \rho\rangle \!\rangle _{q,\omega} = \frac{1}{\pi} K_q \widetilde v_q q^{2} /((\omega+i0)^{2}-q^{2} \widetilde v_q^{2})$, and 
% similar to that of (\ref{eq:rrNoInt})  \cite{Imambekov2012}
%\begin{eqnarray}\label{eq:rrLinearInteract}
% \mbox{Im}\langle\!\langle\rho, \rho\rangle \!\rangle^{ {(0)}} =-\frac{ K_q \, q}{2 \widetilde v_q}\left[\delta\left(q-\omega / \widetilde v_q\right)-\delta\left(q+\omega / \widetilde v_q\right)\right],
% \end{eqnarray}
\begin{equation}\label{eq:rrLinearInteract}
{\cal D}_{q,\omega} ^{ {(0)}} = \tfrac12 { K_q \, q}\left[\delta\left(q\widetilde v_q-\omega \right)-\delta\left(q \widetilde v_q+\omega  \right)\right],
\end{equation}
with the Luttinger parameter, $K_q=1/\sqrt{1+g_q/(\pi v_F)}$,  the group velocity is $\widetilde v_q=v_F/K_q$, and $g_q=\int g(x) e^{-  i  q x} dx$. In this notation, the above $v$   corresponds to $\widetilde v_{q=0}$. 

Let us   provide here the expression for conductivity in the clean Luttinger liquid {\color{black}  (see also Appendix \ref{AppOPE})}. Using the Kubo formula 
$\sigma(\omega)= -\frac{e^2} {\hbar\, \omega} \mbox{Im}\langle\!\langle  j ,\,   j \rangle\!\rangle |_{q\to 0} $ 
and continuity  equation, we have 
\[
	\sigma^{(0)}(\omega)=  {e^2}   K_{0} \widetilde v_{0} \delta({\hbar}\omega)  
	 = {e^2}   v_{F} \delta({\hbar}\omega)\,, 
\]
which shows that % the optical sum, $\int d\omega\, \sigma^{(0)}(\omega)$,  
$\sigma^{(0)}(\omega)$ is not affected by interaction, $K_{q}\neq1$. 

% \subsubsection{Two-boson contribution to the correlator}

Let us discuss ${\cal D}_{q,\omega} ^{ {(1)}} $ for the interaction of finite range, $a$. In the Hamiltonian
\eqref{eq:HamilDens} we  set $H_{U} =0$,  % leave only $H_{0}+H_{int}$,
%\begin{eqnarray}	\label{eq:HamilMain}	H_0= &&\pi v_F \int dx (R(x)^2+L(x)^2)+(4 \pi^2/6m)\int(R(x)^3+L(x)^3)
%	\nonumber\\	&& +(1/2)\int dx_1 dx_2 g(x_1-x_2)\rho(x_1) \rho(x_2), \end{eqnarray}
assume the curvature $1/m$ term small, and find then corrections to the correlator in the lowest order in $1/m$.
Performing  Bogoliubov transformation 
\begin{equation}
\label{eq:Bogoliubov}
\begin{aligned}
R_{q}&=\cosh \theta_{q} \tilde R_q-\sinh\theta_{q} \tilde L_q\\
L_{q}&=\cosh\theta_{q} \tilde L_q-\sinh\theta_{q} \tilde R_q 
\end{aligned}
\end{equation}
with $e^{-2\theta_{q}} = K_q$ or, equivalently, $\tanh2\theta_q=\frac{g_q}{g_q+2 \pi v_F}$,   
one obtains
\begin{equation}
\label{eq:hamilBozed}	
\begin{aligned}
H_0=&(\pi/l) \sum_q \widetilde v_q (\tilde R_q \tilde R_{-q}+ \tilde L_q \tilde L_{-q}) \\
&+ \frac{2 \pi^2 }{3m l^2} \sum_{q_i} [ \Gamma^{(1)}_{q_1,q_2,q_3}(\tilde R_1 \tilde R_2 \tilde R_3+\tilde L_1 \tilde L_2 \tilde L_3) \\
&+3\Gamma^{(2)}_{q_1,q_2;q_3}(\tilde R_1 \tilde R_2 \tilde L_3+\tilde L_1 \tilde L_2 \tilde R_3) ] ,
\end{aligned}
\end{equation}
with $q_1+ q_2 + q_3=0$ and 
\begin{equation}
\begin{aligned}
\Gamma^{(1)}_{q_1,q_2,q_3} &=\cosh\theta_{ q_1}\cosh\theta_{ q_2}\cosh\theta_{ q_3} \\ & -\sinh\theta_{ q_1} \sinh\theta_{ q_2} \sinh\theta_{ q_3}, \\
\Gamma^{(2)}_{q_1,q_2;q_3} &=\sinh\theta_{ q_1}\sinh\theta_{ q_2}\cosh\theta_{ q_3} \\ & -\cosh\theta_{ q_1} \cosh\theta_{ q_2} \sinh\theta_{ q_3}
\end{aligned}
\end{equation}
% and  $\tanh2\theta_q=\frac{g_q}{g_q+2 \pi v_F}$
and $l$ is the size of the system  with the periodic boundary conditions;  one should put $q_{3} = -q_{1}-q_{2}$ in  Eqs.  \eqref{eq:hamilBozed}. Since $g_q$ is even function of $q$, the quantity  $\theta_q$ is also even in $q$. It means that  $\Gamma^{(1)}$ ,  $\Gamma^{(2)}$ are even functions of their arguments, $q_i$.   

The first line of (\ref{eq:hamilBozed}) is the standard Hamiltonian of the Tomonaga-Luttinger model, the second and third lines correspond to the nonlinearity of the spectrum, and they will be considered as the perturbation. 
%Our goal is to find the retarded Green's function (its imaginary part), which can be easily found from the Matsubara Green's function by substituting $ i  \omega_n\rightarrow\omega+ i  0$.

%%%%%%%%%%%%%%%%%%%%%
\begin{figure}
\centering
\includegraphics[width=0.7\linewidth]{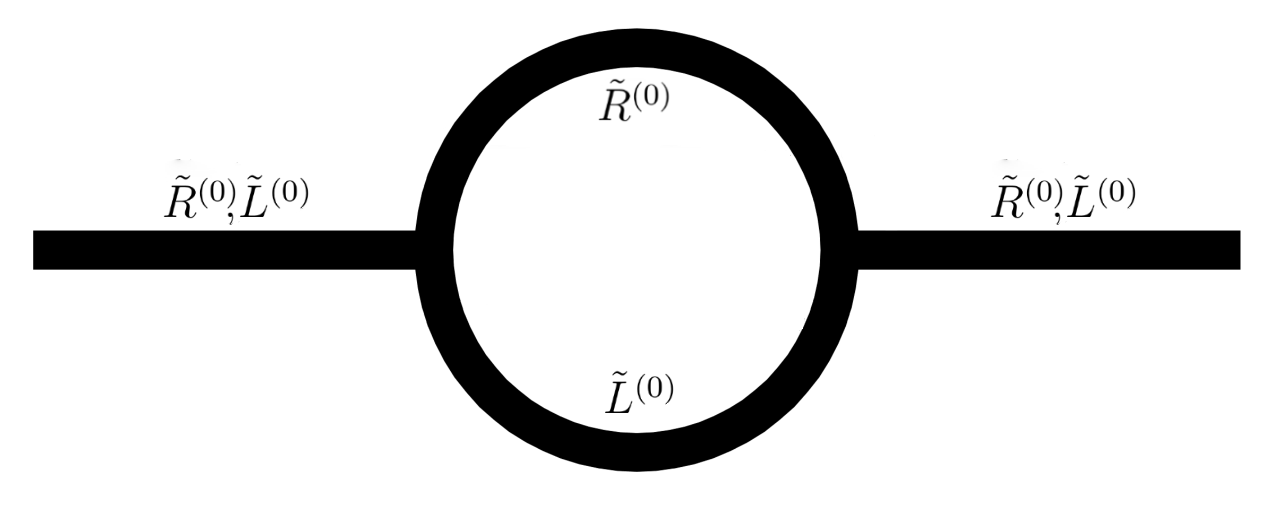}
\caption{Boson diagram of the type $ \langle\tilde R \tilde R \rangle$, $ \langle\tilde R \tilde L \rangle$ etc. It corresponds to the lowest-order correction to the Green's function in $1/m^{2}$, %calculated in App. \ref{app:2bosCont}, and  
describing the boson decaying into two bosons in the intermediate state. As explained in the main text, it is sufficient to use the form of Green's functions with linearized dispersion in the actual calculation. 
% {\color{black}  The symbols $\tilde R^{(0)}$ and $\tilde L^{(0)}$ stand for linearized dispersion operators, when $1/m\rightarrow0$.}   
} 
\label{fig:bosondiagramm}
\end{figure}
%%%%%%%%%%%%%%%%%%%%%

The ordinary correlator is expressed  via the transformed operators $\tilde R$ and $\tilde L$ as  

\begin{equation}
\begin{aligned}
\langle\!\langle\rho, \rho\rangle \!\rangle & =(\cosh\theta_q-\sinh\theta_q)^2\langle\!\langle\tilde\rho, \tilde\rho\rangle \!\rangle \\ 
&= K_{q}\langle\!\langle (\tilde R+\tilde L),  (\tilde R+\tilde L) \rangle \!\rangle \,,
\end{aligned}
\end{equation}
and we come to  Eq. \eqref{eq:rrLinearInteract} above.

We find the correction to the Green's function in the lowest (second) order in $1/m$. As explained above, it amounts to calculation of   diagrams shown  in Fig \ref{fig:bosondiagramm}. A similar calculation was done in \cite{aristov2007}, and we  generalize it here to the case of non-point-like interaction between  particles. Our goal is to find the imaginary part $ {\cal D}_{q,\omega}^{(1)}$ away from the pole, in which case the diagrams remain finite, with no further regularization required. As was noted in \cite{aristov2007}, 
the terms with the vertex $\Gamma^{(1)}$  do not affect the second order perturbation of the Green's function imaginary part $ {\cal D}_{q,\omega}^{ {(1)}}$.

Second order corrections in $m^{-1}$,  defining $ {\cal D}_{q,\omega}^{ {(1)}} $, stem from the vertex   
$\Gamma^{(2)}$, which is non-zero in presence of interaction, $g_{q}\neq 0$.
The corresponding contributions are given by  expressions
\begin{equation}\begin{aligned}
\mbox{Im} \langle\!\langle \tilde R\tilde R \rangle \!\rangle_{q,\omega}^{{(1)}} & = K_q^{-1}{\mathcal C} (q,\omega)  \frac{ \left(\Gamma^{(2)}_{q,q';q'-q} \right)^2}{(q \tilde{v}-\omega)^2} \,, \\ 
\mbox{Im} \langle\!\langle  \tilde L\tilde L \rangle \!\rangle_{q,\omega}^{\text{(1)}}&= K_q^{-1}{\mathcal C} (q,\omega) \frac{  \left(\Gamma^{(2)}_{q,q'-q;q'} \right)^2}{(q \tilde{v}+\omega)^2} \,, \\ 
\mbox{Im} \langle\!\langle  \tilde R\tilde L \rangle \!\rangle_{q,\omega}^{\text{(1)}} & =K_q^{-1} {\mathcal C} (q,\omega) \frac{  \Gamma^{(2)}_{q,q';q'-q}\Gamma^{(2)}_{q,q'-q;q'}}{(q \tilde{v}-\omega)(q \tilde{v}+\omega)} \, , \\
\mbox{Im} \langle\!\langle  \tilde L\tilde R \rangle \!\rangle_{q,\omega}^{\text{(1)}}& =\mbox{Im} \langle\!\langle  \tilde R\tilde L \rangle \!\rangle_{q,\omega}^{ {(1)}} \,, 
\end{aligned}
\end{equation}
where $q'=\frac{\omega}{2 \tilde{v}}+\frac{q}{2}$.
The total  spectral density correction is then 
\begin{equation}\begin{aligned}
\label{eq:GeneralCorrel}
{\cal D}_{q,\omega}^{(1)} & = 
%	\frac{ \hbar^2 q^2(\omega^2-(q \tilde{v})^2)}{  m^2 2^5 \tilde{v}^3} \frac{\sinh(\hbar \omega/2T)}{\sinh(\hbar\,[\omega+q \tilde{v}]/4T)\sinh(\hbar\,[\omega-q \tilde{v}]/4T) } \nonumber \\	&\times
{\mathcal C} (q,\omega)
\left[ \frac{\Gamma^{(2)}_{q,\frac{\omega}{2 \tilde{v}}-\frac{q}{2};\frac{\omega}{2 \tilde{v}}+\frac{q}{2}}}{q \tilde{v}+\omega}+\frac{ \Gamma^{(2)}_{q,\frac{\omega}{2 \tilde{v}}+\frac{q}{2};\frac{\omega}{2 \tilde{v}}-\frac{q}{2}}}{q \tilde{v}-\omega}\right] ^2 \,, \\
{\mathcal C} (q,\omega) & =  K_q \frac{  q^2}{32 m^2   \tilde{v}^3} \frac{(\omega^2 - (q \tilde{v})^2) \sinh(\frac{\omega}{2T})}{\sinh(\frac{\omega + q\tilde{v}}{4T}) \sinh(\frac{\omega - q\tilde{v}}{4T})} \,,  	 
\end{aligned}
\end{equation}
% \[ \max \min   \mbox{italic} \]
for $q$ away from $\pm\omega/\tilde{v}$.
%Note that the conductivity is determined by the integral of  (\ref{eq:GeneralCorrel}) over the area away from its  poles $\omega=\pm q \tilde{v}$ (refer to (\ref{eq:AwayPole}). 

The analog of the  formula \eqref{eq:GeneralCorrel} for a point-like interaction was found in \cite{aristov2007}. The point-like interaction corresponds to {\color{black}  formally letting $a\rightarrow0$ (while keeping $\lambda_{F}\ll a$, see  Appendix \ref{AppOPE})}, which results in $(\Gamma^{(2)}_{q,\frac{\omega}{2 \tilde{v}}+\frac{q}{2} ;\frac{\omega}{2 \tilde{v}}-\frac{q}{2}})^2 \simeq (\Gamma^{(2)}_{0,0;0})^2=   (K-1)^2/16K$ with  $K=K_{q=0}$. The Eq.\    \eqref{eq:GeneralCorrel}  reduces then to Eq.\ (51) in \cite{aristov2007}.

\section{\label{sec:inter} The conductivity of  interacting 1D liquid }

% \subsection{One-boson contribution to the correlator}

Let us first calculate one-boson contribution which means using  expression   (\ref{eq:rrLinearInteract}) in Eq.\  (\ref{eq:DmitSigmaGENERAL}). As explained in the previous section, when condition $1/d\gg q_+-q_-$ is satisfied, the fine structure of the peak can be neglected and substituted by delta function of the linearized dispersion limit,  $1/m\rightarrow0$. Then we return to conditions 
\eqref{eq:LargeScale}    
with $v_F$ replaced by renormalized $\tilde{v}$. 
 %  This yields	 \begin{eqnarray}   	\label{eq:LargeScale}
 % 	\omega\ll \frac{\tilde{v}}{d} \frac{m \tilde{v}^2}{\max (T,\omega)}  \end{eqnarray}	 
 % with the renormalized $\tilde{v}$.

The expression (\ref{eq:rrLinearInteract}) contains $q=\omega / \widetilde v_q$ with  $ \widetilde v_q$ depends on $q$ itself. 
After simple calculation we find 
% To simplify the final result, we additionally demand % apply some weak condition. Specifically, we use 
% $\omega / v_{F} \ll  1/a$, which leads to % then, after using (\ref{eq:DmitSigmaGENERAL}) we obtain 
\begin{equation}
\label{eq:sig1Inter}
\sigma_{1}(\omega)=
\frac{e^{2} K_{\omega / v}  {W}_{\omega / v}}{2 \pi \hbar m^{2} {\color{black}  K_{0}^{4} } \widetilde v_{\omega / v}^{4}}
\end{equation} %	{\color{black}  -- repeat}
This result is  similar to the non-interacting case (\ref{eq:NonInteracResult}), except for certain renormalization. 
We see that this one-boson contribution does not depend on temperature when inequality (\ref{eq:LargeScale}) holds.
Since we assume that ${W}_q$ decreases faster than any power with $q$ beyond $1/d$, we see that $\sigma_{1}(\omega)$ is suppressed at $|\omega| \agt \omega_d$ where the scale $$\omega_d = \tilde{v} /d \,.  $$

%As noted in the Introduction, the problem of 

% \subsection{Two-boson contribution to the correlator}

%We integrate {\color{black}  these two contributions $\mbox{Im}\left\langle\rho \rho\right\rangle_{q,\omega}^{ {(1)}}$ and $\mbox{Im}\left\langle\rho \rho\right\rangle_{q,\omega}^{ {(1)}}$} separately.
We calculate next the  two-boson contribution, focusing on the long tails in density correlator in Fig. \ref{fig:ImRR}.  If $T\gg \omega_d$ or $\omega\gg \omega_d$, then the area near $q\approx1/d$ of the long tails contributes most significantly to the conductivity.  However, this approach meets difficulties when $\omega \approx \omega_d$, because it is the peak rather than  the long tails, which may mostly contribute in this regime, see Fig.\ \ref{fig:ImRR}.
%Nevertheless, we find no singularities for this set of parameters. 
Specifically, we should require $|\omega-\omega_d|\gg q_+-q_-$, or
% {\color{black}  explain - done, cf. def after Eq. \eqref{eq:rrNoIntNonLinear} }
\begin{equation}
\label{eq:AwayPole}
\left |\omega-\omega_d \right |\gg \frac{\omega_d \max(T,\omega_d)}{ m v_F^2},
\end{equation}
for this simple estimate we let $\omega\approx\omega_d$ in \eqref{eq:qplusqminus}.    We may conclude that focusing on the tails of density correlator for calculation of the conductivity in presence of interaction is questionable at $\omega  \approx \omega_d$. In particular, our Eq. \eqref{eq:vdwT} below contains a mild (logarithmic) singularity. 

The general expression \eqref{eq:DmitSigmaGENERAL} with correlator \eqref{eq:GeneralCorrel} can be brought to compact form in particular cases considered below. When integrating in \eqref{eq:DmitSigmaGENERAL}, we take into account the fact  that \eqref{eq:GeneralCorrel} is obtained for $q$ away from the apparent poles at  $\pm\,\omega/\tilde{v}$. \footnote{Generally these apparent poles may be regularized by shifting them in the complex plane $\omega\rightarrow\omega+ i \delta$ and then taking the real part of the result. However for approximate calculation we may use, e.g.,  the following replacement:  $ {1}/{(q \tilde{v}-\omega)}\simeq -{1}/{\omega}$ for $q \tilde{v}\ll \omega$.}

The results for the higher temperature regime, $T\gg \omega_d $, are summarized as follows 
\begin{equation}
\begin{aligned}
\sigma_{ {2}}(\omega) &=\frac{ e^2 T {\color{black}  K_{0}^{-4} } }{8 m^4 \tilde{v}^5 \omega^{2}}\int_{0}^{\infty} q^{2}   {W}_q K_q \left[   2\Gamma^{(2)}_{q,\frac{q}{2};\frac{q}{2}}\right]^2 \frac{d q}{2 \pi}   \,,\\ 
&   \simeq 
% \frac{ \, T \, \Xi_0  }{12  \, \tilde{v}^5\, \omega^{2}d^3} \,,\quad  (T\gg \tilde{v}/d\gg \omega) \,.
\frac{  T \, \omega_d^{3} \, \Xi_0}{12  \,  \omega^{2} } \,,\quad  (T\gg \omega_d\gg \omega) \,.
\end{aligned} \label{eq:Tvdw}	
\end{equation} 
\begin{equation}
\begin{aligned}
\sigma_{ {2}}(\omega) & =\frac{ e^2 T {\color{black}  K_{0}^{-4} }}{8 m^4 \tilde{v} \omega^{6}}\int_{0}^{\infty} q^{6}   {W}_q K_q  \left[    2\Gamma^{(2)}_{q,\frac{\omega}{2 \tilde{v}};\frac{\omega}{2 \tilde{v}}}\right]^2 \frac{d q}{2 \pi} 
\,, \\ &  \simeq 
% \frac{    T \, \Xi_\omega  }{7  \, \tilde{v}\, \omega^{6}d^7} \,,\quad (T\gg \omega\gg \tilde{v}/d) \,. 
\frac{    T \, \omega_d^7\, \Xi_\omega  }{7  \,   \omega^{6} } \,,\quad (T\gg \omega\gg \omega_d) \,. 
\end{aligned}\label{eq:Twvd}		
\end{equation} 
\begin{equation}
\begin{aligned}
\sigma_{ {2}}(\omega)&=\frac{  \hbar \, e^2 {\color{black}  K_{0}^{-4} }}{32 m^4  \tilde{v} |\omega|^{5}}\int_{0}^{\infty} q^{6}   {W}_q K_q   \left[   2 \Gamma^{(2)}_{q,\frac{\omega}{2 \tilde{v}};\frac{\omega}{2 \tilde{v}}}\right]^2 \frac{d q}{2 \pi}   
\,, \\ &  \simeq 
% \frac{\hbar \, \Xi_\omega  }{28 \,  \tilde{v} \,|\omega|^{5}d^7}\,, \quad (\omega\gg T\gg \tilde{v}/d) \,, 
\frac{\hbar \,   \omega_d^7 \, \Xi_\omega }{28 \,   |\omega|^{5} }\,, \quad (\omega\gg T\gg \omega_d) \,, 
\end{aligned}\label{eq:wTvd}			
\end{equation} 
where we defined  
% $$\Xi_\omega=\frac{ e^2 \omega_d^2 K_0 {W}_0 \,\big( \Gamma^{(2)}_{0,\omega/2\tilde{v};\omega/2\tilde{v}}\big)^2}{ \pi  (m  \tilde{v}^2)^4}$$
$$\Xi_\omega=\frac{ e^2 %  K_0 
{\color{black}  K_{0}^{-3} } {W}_0 }{ \pi  (m  \tilde{v}^2)^4}\,\big( \Gamma^{(2)}_{0,\omega/2\tilde{v};\omega/2\tilde{v}}\big)^2\, .$$

Similarly, in the lower temperature regime, $T \ll \omega_d$, we obtain 
% \begin{eqnarray} \label{eq:vdTw}	
% \sigma_{\text{2}}(\omega)=\frac{ 12 \cdot 28\, \Xi \, T^4 }{\hbar^3  \tilde{v}^8 \omega^{2}  } \,, 
% \end{eqnarray}  for $(\tilde{v}/d\gg T\gg \omega)$, 
% \begin{eqnarray} \label{eq:vdwT}	
% \sigma_{\text{2}}(\omega)\approx	 \frac{\hbar \, \Xi \, \omega^2}{ \tilde{v}^8} \,,  \end{eqnarray}
% for $\quad (\tilde{v}/d\gg \omega\gg T)$, and  \begin{eqnarray} \label{eq:wvdT}	
% \sigma_{\text{2}}(\omega)\approx  \frac{ \hbar \,\Xi }{16\,   \tilde{v}\, \omega^{5}d^7},  \end{eqnarray}
% for $\quad(\omega\gg \tilde{v}/d\gg T)$. 

% \begin{equation}
% \begin{aligned}
% 	\label{eq:vdTw}	
% 	\sigma_{\text{2}}(\omega) & \approx \frac{ 12 \cdot 28\, \Xi \, T^4 }{\hbar^3  \tilde{v}^8 \omega^{2}  } \,, \quad 
% 	\tilde{v}/d\gg T\gg \omega  \, \\ 
% 	% \label{eq:vdwT}	
% 	& \approx	 \frac{\hbar \, \Xi \, \omega^2}{ \tilde{v}^8} \,, 
% 	\quad   \tilde{v}/d\gg \omega\gg T   \,, \\  
% 	% \label{eq:wvdT}
% 	& \approx  \frac{ \hbar \,\Xi }{16\,   \tilde{v}\, \omega^{5}d^7} \, , 
% 	\quad \omega\gg \tilde{v}/d\gg T \, . 
% \end{aligned}
% \end{equation}	 
 
\begin{numcases}{\sigma_{ {2}}(\omega) \approx }
%\frac{ 32 \pi^4\, \Xi_0 \, T^4 }{ 15 \hbar ^3 \tilde{v}^8 \omega^{2}  }  
\frac{ 32 \pi^4 \, T^4\, \Xi_0}{ 15 \hbar ^3   \,\omega^{2}  }  
\,, &  \text{$\omega_d \gg T\gg \omega $}  \label{eq:vdTw} \\
% \frac{\hbar \, \Xi_0 \, \omega^2}{ 4\tilde{v}^8} 
\frac{\hbar  \, \omega^2\, \Xi_0}{ 4 \,} 
\,, & \text{$\omega_d \gg \omega\gg T$} \label{eq:vdwT} \\
% \frac{ \hbar \,\Xi_\omega}{28\,   \tilde{v}\, |\omega|^{5}d^7} 
\frac{ \hbar \,  \omega_d^7\,\Xi_\omega}{28\,   |\omega|^{5} } 
\,, & \text{$\omega\gg \omega_d \gg T$} \label{eq:wvdT}
\end{numcases}

Notice that \eqref{eq:wTvd} and \eqref{eq:wvdT} are equal, although in different regimes.   The results \eqref{eq:wTvd}, \eqref{eq:vdwT}  and \eqref{eq:wvdT} correspond to the low temperature limit, $T \rightarrow 0$, when the temperature-dependent factor in \eqref{eq:GeneralCorrel} becomes a step function equal to zero at $\omega<q \tilde{v}$, cf. \cite{pustilnik2006dynamic}. 
% The integration in \eqref{eq:vdTw} and \eqref{eq:vdwT} is truncated at $q\approx 1/l_T$ and otherwise at $q\approx 1/d$.
{\color{black}  The small factor $q^2$ in integrands \eqref{eq:Tvdw}-\eqref{eq:Twvd}  is inherited
from Eq. \eqref{eq:DmitSigmaGENERAL}, and extra powers of $q$ come from the phase-space and vertex factors.}

The simpler formulas above with the quantity $\Xi_\omega$ were obtained keeping in mind that $1/d \ll 1/a$, i.e., the range of interaction is smaller than the scale of random potential. For weak interaction, $g_0/\pi v_F\ll 1$, the formula for $\Xi_\omega$ is additionally simplified with % $\Gamma^{(2)}_{0,0;0} \to  g_{0}/2\pi v_F$ and 
$\Gamma^{(2)}_{0,\frac{\omega}{2 \tilde{v}};\frac{\omega}{2 \tilde{v}}} \to  g_{\frac{\omega}{2 \tilde{v}}}/2\pi v_F$, we also have $K_0 \to 1$ and $\tilde{v} \to v_F$.  Notice that   $\sigma_{ {2}}(\omega\sim \omega_d)$ at $T\sim \omega_d$ is small in comparison with $\sigma_1(0)$ thanks to the factor  $\Xi_0$, which contains additional smallness both in interaction and disorder. 

%smoothness of disorder,  $\sigma_{ {2}}(\omega\sim \omega_d) /\sigma_1(0) \sim (\omega_d /E_F)^2 ( g_0/v_F)^2 $, with $E_F=m v_F^2/2$. 

\begin{figure}
\centering
\includegraphics[width=0.5\linewidth]{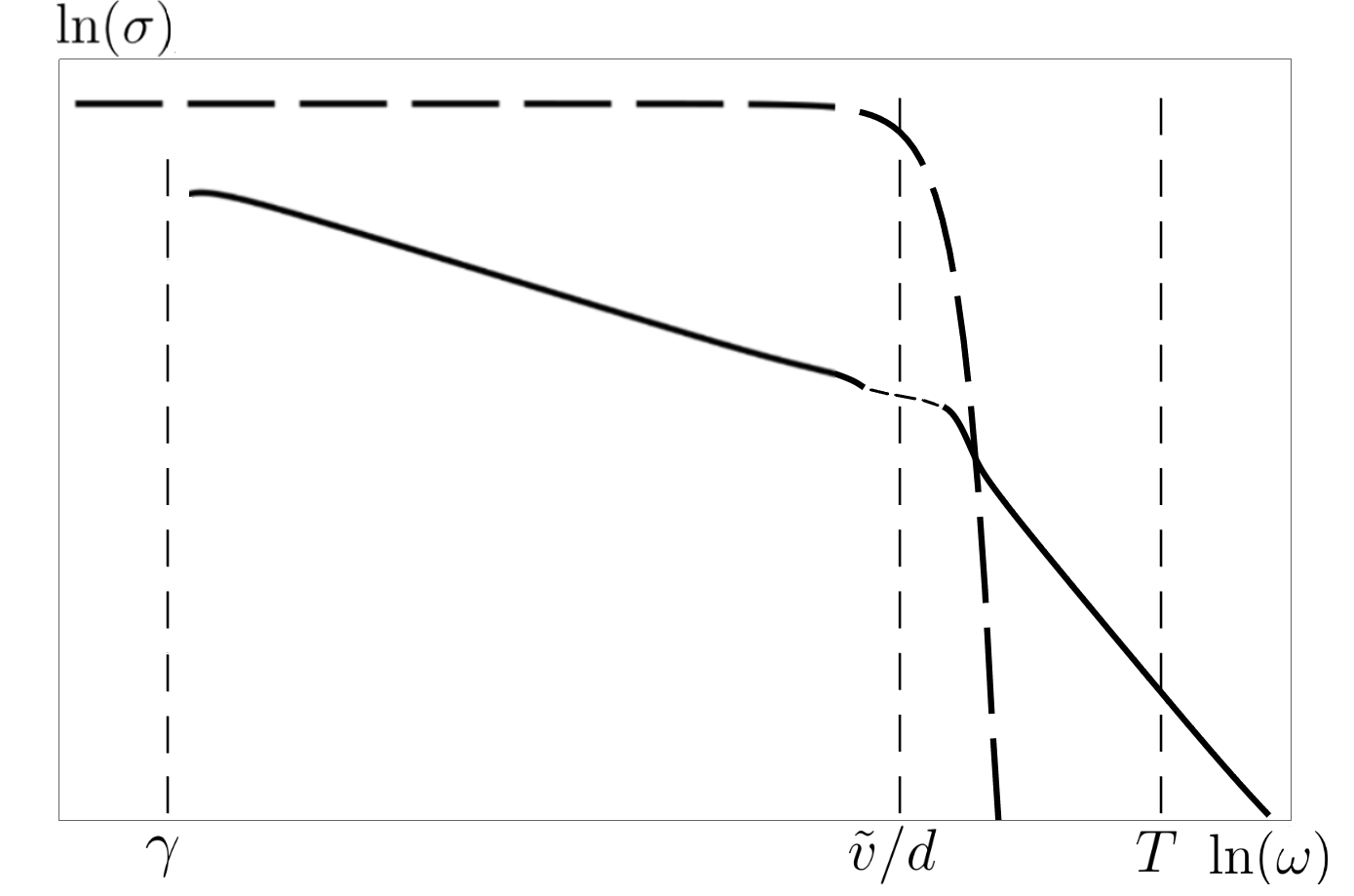}
\caption{ % {\color{black}  Redraw the figure for smaller frequencies, showing the scale $\gamma$ and saturation of the lorentzian. } {\color{black}  Done!} 
One-boson $\sigma_1$ (dashed line) and two-boson $\sigma_2$ (solid line) conductivities as a function of $\omega$ in log-log format. The temperature is chosen high $T\gg \omega_d$.  Conductivity $\sigma_2$ decreases as a power of $1/\omega$, while $\sigma_1$ does as $\mathrm{W}_{\omega / v_{F}}$. Since $\mathrm{W}_{q}$ dependence is exponential, one can see that there are two areas of frequencies where $\sigma_2>\sigma_1$ and both of them are valid within boson approximation. Note that there is a area of $\omega$ in $\sigma_2$ near point $\tilde{v}/d$, that our method cannot describe, see condition of applicability (\ref{eq:AwayPole}).}
\label{fig:sigmas12}
\end{figure}

The total conductivity at finite frequencies consists of two contributions $\sigma_1$ and $\sigma_2$ 
% One of them is dominant, depending on the set of parameters, which means that 
and none of these contributions can be ignored, see Fig. \ref{fig:sigmas12}.
The part  $\sigma_2$ is dominant at higher frequencies, $\omega \agt \omega_d$. 
%  and also in the low frequency domain, 
% \begin{equation}
%     \omega \alt \omega_1 \equiv  \frac{g_0}{2\pi v_F} \frac{ \min [\, \omega_d^{3/2} T^{1/2},\, T^2  ] }{E_F}\,.
 %   \label{def:omega1}. \end{equation}    
 
In the regime of  small energies, $\omega \ll T$, Eqs.\ \eqref{eq:Tvdw}, \eqref{eq:Twvd}, \eqref{eq:vdTw}, the part $\sigma_2$ is proportional to $T$, which allows us to identify $\sigma_2$ as {\it inelastic} contribution to conductivity. By contrast, $\sigma_1$ is independent of temperature, and it is natural to refer to it as {\it elastic} contribution to $\sigma(\omega)$. 

{\color{black}  The elastic contribution to the conductivity $\sigma_{1}(\omega)$ may be viewed as the contribution of a single quasiparticle (bosonic in the Luttinger liquid) moving in a random potential under an AC electric field. We saw in the classical case \eqref{eq:ClassResult} that even one noninteracting particle gives a contribution to the conductivity. In the interacting case \eqref{eq:sig1Inter} the parameters of  this quasiparticle are slightly renormalized, so that the elastic contribution $\sigma_{1}(\omega)$ depends only weakly on interaction and temperature. The inelastic contribution $\sigma_{2}(\omega)$, on the other hand, comes from processes in which two quasiparticles scatter both from each other and from the random potential. This process is explicitly proportional to the strength of interaction at weak coupling, $\sim \Xi_0 \propto g^2$, and depends strongly on  temperature. 
We stress that both elastic and inelastic contributions vanish in the absence of the curvature. 
}

Let us now address one particular case, when our results can be checked by the alternative method of kinetic equation.  
This case 
requires a necessary condition $\omega\ll \omega_d $ for the method to be applied. Other requirements include $\omega\ll T$ and the weakness of interaction.
This leaves us with two options: Eq.  (\ref{eq:Tvdw}) and Eq. (\ref{eq:vdTw}). For example, in particular case $T\gg \omega_d$, the conductivity is given by (\ref{eq:Tvdw}), which simplifies to
\begin{eqnarray}
\label{eq:sigmaboz}
\sigma _2(\omega)=\frac{e^2 T {\color{black}  K_{0}^{-4} }}{16 \pi m^4 v_F^5 \omega^2} \int_0^\infty dq\, q^2 W_q \left( \frac{g_{q/2}}{\pi v_F}\right) ^2.
\end{eqnarray}
This formula for inelastic part of conductivity  coincides with the one obtained by using the kinetic equation.  If the interaction is weak and short-ranged, that approach gives a general formula that works for any ratio of $T$ to $\omega_d$. 
\[
\sigma _2(\omega)= \frac{e^2{\color{black}  K_{0}^{-4} }}{256\pi m^4 v_F^3 T  \omega^2}\left( \frac{g_{0}}{\pi v_F}\right) ^2
\int_{0}^{\infty} dq\;
\frac{q^{4}}{\sinh^{2}\!\left(\dfrac{v_{F}\,q}{4T}\right)}
W_{q}
\]
In the special case that matches our earlier limits, this general result becomes Eq. \eqref{eq:vdTw}.  (This general result can also be derived from the bosonization expression \eqref{eq:GeneralCorrel}.)	

\subsection{\label{subsec:lowFreq} The case of low frequency} 

Inspecting the inelastic contribution to conductivity at the lowest frequencies,  Eqs. \eqref{eq:Tvdw}, \eqref{eq:vdTw}, we see a 
non-integrable divergence, $\sigma \sim \omega^{-2}$, which violates the so-called optical sum rule \cite{Aristov2002,Aristov2005}. The sum rule states that the integral, $\int d\omega\, \sigma(\omega)$, is proportional to the density of states at the Fermi level and is not affected  by interactions.  
It shows that the use of  the formula (\ref{eq:DmitSigmaGENERAL}) leads to unsatisfactory results in the limit of small frequencies, when 
interaction and weak disorder are simultaneously present.

% It turns out that  the limit $\omega\to 0$ requires  consideration of higher orders of the correlator $W(x)$. 
A possibility to proceed with the limit $\omega\to 0$ in such cases was proposed   in \cite{wolfle1972homogeneous}. 
It was shown that the complex-valued conductivity can be  represented by the expression
\begin{equation}
	\label{eq:SigmageneralThroughM}
	\sigma_{complex} 
	(z)=\frac{  e^2 v_F}{\pi \hbar} \frac{i}{z+M(z) }\,,
\end{equation}
where $z=\omega+i 0 $ and $M(z)= M'(z) + i M''(z) $ is the  complex-valued memory function.  
%The real part of the memory function, $M'(\omega) $, determines the change 
%of the weight before delta function contribution and/or the change 
%$v_F \to \tilde{v}$. 

Let us for a moment assume that $M'(\omega) $ is small in  case of weak disorder and small interaction. Then we neglect it and  
the real part of the conductivity at $\omega \neq 0$ reads 
\begin{equation}
	\label{eq:mkvPLUSwkv}
	\sigma(\omega)=\frac{ e^2 v_F}{\pi \hbar} \frac{M''(\omega)}{\omega^2+( M''(\omega) )^2} \,. 
\end{equation}
In the first order of  random potential correlator the imaginary part, $M''(\omega) $, according to  \cite{wolfle1972homogeneous}, is given by 
\[ M''(\omega)=\frac{ \pi}{v_F m^2 \omega}\int  \frac{d q}{2 \pi} q^{2}  {W}_q {\cal D}_{q,\omega} \,. \] 
%Comparing this expression with Eq.\  %(\ref{eq:DmitSigmaGENERAL}), we have
%\begin{equation}
%\begin{aligned}
%\label{eq:MtwoSigmas}
%M''(\omega) & = \frac{\pi \hbar \omega^2}{e^2 {v}_F} \left[ %\sigma_1(\omega) + \sigma_2(\omega) \right] 
%\,, \\ & = M''_1(\omega) + M''_2(\omega)  \,, 
%\end{aligned}
%\end{equation}
%which means that  Eq.\  \eqref{eq:DmitSigmaGENERAL} should be %viewed as the leading large-$\omega$ asymptotics for the %conductivity. 

% In the leading order, i.e. at $\omega \gg M''(\omega)$, Eq.(\ref{eq:mkvPLUSwkv}), reduces to Eq.(\ref{eq:DmitSigmaGENERAL}). 
% %, thereby confirming the latter. 
% We have already identified the leading correction in the $1/\omega$ expansion—proportional to $1/\omega^{2}$ and given in Eq.\  \eqref{eq:vdTw}.
% At lower frequencies, however, Eq.\  \eqref{eq:mkvPLUSwkv} no longer describes the conductivity of our system.   
%  Below we derive the next term in the series, yielding a more accurate low‑frequency conductivity:

The leading asymptote of  Eq.\ \eqref{eq:mkvPLUSwkv}, corresponds to $\omega \gg M''(\omega)$ and reduces to Eq.\ \eqref{eq:DmitSigmaGENERAL}. We identify this  $1/\omega^{2}$ regime with the above  Eq.\ \eqref{eq:vdTw}. 
We show below, however, that Eq.\ \eqref{eq:mkvPLUSwkv} badly describes the conductivity at lower frequencies.
Namely we extract the next term in the asymptotic expansion of $\sigma(\omega)$ in powers of $1/\omega$, and find     
\begin{equation} 
\operatorname{Re}[\sigma_{\text{low}}(\omega\neq 0)]\simeq \frac{e^{2}v_{F}}{\pi\hbar}
\left(\frac{A}{\omega^{2}}-\frac{B}{\omega^{4}}\right) \,. 
\label{sigma-expa0} \end{equation}
% where the coefficients $A$ and $B$ do not satisfy the constraints implied by Eq.(\ref{eq:SigmageneralThroughM}). 
If Eq.\ \eqref{eq:mkvPLUSwkv} would hold, then one would have   $A =  M''( 0)$, and  $B=  [M''( 0)]^{3}$. We will see that the ratio $B/A^3 \neq 1$, which particularly may mean that our assumption about smallness of $M'(\omega)$, leading to \eqref{eq:mkvPLUSwkv}, was incorrect. 

%a condition that, as we show below, is not fulfilled. These constraints cannot be met in any other way.To see this, temporarily ignore the real part of $M(\omega)$ and expand Eq.(\ref{eq:mkvPLUSwkv}) in powers of $\omega$.  

Equation \eqref{sigma-expa0} defines the applicability condition of the above expressions \eqref{eq:vdTw} and \eqref{eq:Tvdw}, in the form
% These formulas hold until the successive terms in Eq. \eqref{sigma-expa0} become comparable, which occurs when  
\[
\omega \agt \omega_{\text{min}}
      \equiv \sqrt{{B}/{A}}\, .
\]
Below we will refer to this characteristic scale as $\omega_{\text{min}}=\gamma$.

Representing  $A$ and $B$ as $A= \zeta\gamma$ and  $B =  \zeta\gamma^{3}$  we may write  
 the expansion for small $\omega\ll \min[T,\omega_{d}] $
\begin{equation}
	Re\, \sigma_{low}
	(\omega\neq0) \simeq\frac{  e^2 v_F}{\pi \hbar} \left(  f(\omega) + \frac{ \zeta\gamma}{\omega ^{2}}  - \frac{ \zeta\gamma^{3}}{\omega ^{4}} 
	\right) \,,
	\label{sigma-expa}
\end{equation}	
with  the  first two terms here correspond to $\sigma_{1}$ and $\sigma_{2}$, Eqs. \eqref{eq:sig1Inter}  and 
\eqref{eq:Twvd}, \eqref{eq:vdTw}, that is  $f(\omega)$ defined  as $\sigma_{1} = \frac{  e^2 v_F}{\pi \hbar}  f(\omega)$. 

% It shows that if, in addition to contribution $Re\, \sigma \propto \omega^{-2}$, we  find the next term, $Re\, \sigma \propto \omega^{-4}$, then we can restore both parameters $\zeta$ and $\gamma$. Judging only by the  contribution, $\propto x\gamma/ \omega ^{2}$, we are unable to  distinguish between $\zeta$ and $\gamma$.

% $\gamma \ll \min[T,\omega_{d}]$

% In what follows, we propose and substantiate the following picture. 

The form of Eq.\ \eqref{sigma-expa} suggests that   
in the lowest order of random potential and  at  frequencies, $\gamma \ll \omega \alt \min[T,\omega_{d}]$, 
the complex conductivity  may be conveniently represented by three terms, 
\begin{equation}
	\sigma_{low} 
	(\omega)\simeq\frac{  e^2 v_F}{\pi \hbar} \left( \frac{i \zeta_{0} }{\omega+i0} + \frac{i \zeta}{\omega+i\gamma} 
	+ f(\omega) \right) \,,
	\label{sigma-low}
\end{equation}	 
%with $f(\omega)$ defined by Eq. \eqref{eq:sig1Inter} as $\sigma_{1} = \frac{  e^2 v_F}{\pi \hbar}  f(\omega)$. 
where the first term corresponds to the $\delta$-function in real part of the conductivity. 
Performing analytical continuation to the smallest frequency domain, $\omega \alt \gamma$, we expect that Eq. \eqref{sigma-low} holds in the whole region $ | \omega| \alt \min[T,\omega_{d}]$. 
We show below that  $\zeta$ is small for weak   disorder and, together with  the optical sum rule, $\int d\omega\, Re\, \sigma(\omega) =  {  e^2 v_F}/{  \hbar} $, it leads to conclusion that the weight of the $\delta$-function in conductivity, $\zeta_{0}\simeq 1$.

%We argue, however, that formal correctness of the general %expression, Eq.\ \eqref{eq:SigmageneralThroughM}, does not imply %the possibility to discard the real part of the memory function, %$M'(\omega)$. Indeed, one can see that the part, $M''_2(\omega)$, %calculated in the  above manner, is finite at $\omega\to 0$ and %hence it provides the Drude form of conductivity at small %frequencies,   $\sigma(\omega) \sim \tau /(1-i\omega \tau)$. The %latter form satisfies the sum rule, but it means that the part of % $\sigma(\omega)$, proportional to $\delta(\omega)$, is %altogether absent, which statement  in one spatial dimension has %to be  checked.   

%To clarify our subsequent derivation, we adopt the following simplified picture. 

% in  first order of disorder strength consists of three terms: (i) $\delta$-function, (ii) the smooth part, $\sigma_{1}$, surviving even without interaction, and (iii) the part $\sigma_{2}$ appearing due to interaction, of finite weight and  Lorentzian shape, whose $\omega$-dependence is given by \eqref{sigma-expa}. Overall, the optical sum rule is preserved.

\subsection{ Contribution $\omega^{-4}$  to conductivity}

We found the expression \eqref{dJdt} for the uniform dynamic component of the current in the form, 
$\partial_{t} j(x, t)=-\frac{1}{m} \rho(x, t) \partial_{x} U(x)$, 
which leads to $\omega^{-2}$ contribution to $\sigma(\omega)$. We suggest, that $\omega^{-4}$ contribution can be obtained by further differentiating $\partial_{t} j(x, t)$ over time, so that instead of Eq.\ \eqref{eq:Kubo} we write 
 \begin{equation} 
	\sigma(\omega)= -\frac{e^2\,\mbox{Im}\langle\!\langle \overline{\partial^{2}_{t} j ,\, \partial^{2}_{t} j} \rangle\!\rangle |_{q=0}}{\hbar\, \omega^{5}} \,.
\end{equation}
In  the uniform limit, $q=0$, we write $\partial_{t}^{2} j(x, t)=-\frac{{\color{black}  K^{-2} }}{m} \partial_{t}\rho(x, t) \partial_{x} U(x)$, which after using continuity equation and integration by parts leads to   
\begin{equation}
	\begin{aligned}
		\sigma(\omega) & =-\frac{e^2}{\hbar\, m^{2} {\color{black}  K^{4} } \omega^{5}} \mbox{Im}\langle\!\langle  
		\overline{(\partial_{x}^{2} U) j ,(\partial_{x}^{2} U)j } \rangle\!\rangle _{q=0} \,.
	\end{aligned}
\end{equation}
where the current is given by Eq.\ \eqref{eq:CurrentDiff}.  In the above current response function,  $\langle\!\langle   j , j \rangle\!\rangle$,  the linear-in-density components of the current $j = v_{F}(R-L )$ do not lead to the desired term, $\sigma(\omega)\propto \omega^{-4}$. Indeed,  in the leading order % of curvature, disorder and interaction the leading contribution in both $ v_{F}^{2} \mbox{Im}\langle\!\langle   (R-L ) ,   (R-L ) \rangle\!\rangle$ and  $ \pi v_{F}/m \mbox{Im} \langle\!\langle   (R-L ) ,   (R^{2}-L^{2} ) \rangle\!\rangle$ 
the expression
$ v_{F}^{2} \mbox{Im}\langle\!\langle   (R-L ) ,   (R-L ) \rangle\!\rangle$
has the form $\pm \delta(q v_{F}\mp \omega) $. It only gives corrections to above $\sigma_{1}$ and is not interesting here. Consideration of the mixed  terms,  $ \propto \mbox{Im} \langle\!\langle   (R-L ) ,   (R^{2}-L^{2} ) \rangle\!\rangle$, supplied by the leading order corrections in curvature, does not also result in $\omega^{-4}$ contribution to conductivity. 

At the same time, the correlation between quadratic-in-density components of the current is important and  corresponds to expression (cf. Eq. \eqref{eq:DmitSigmaGENERAL})
%\begin{equation*}
%	\begin{aligned}
%		\sigma(\omega) & =-\frac{e^2 \pi^{2}}{\hbar\, m^{4} \omega^{5}} \mbox{Im}\langle\!\langle  
%		\overline{(\partial_{x}^{2} U) (R^{2}-L^{2} ) ,(\partial_{x}^{2} U)(R^{2}-L^{2} ) } \rangle\!\rangle  \,.
%	\end{aligned}
%\end{equation*}
%which results in 
\begin{equation}
	\begin{aligned} 
		\label{eq:omega5}
		\sigma(\omega) & =\frac{e^2 \pi^{2}}{\hbar\, m^{4} {\color{black}  K ^{4} } \omega^{5}} \int  \frac{d q}{2 \pi} q^{4}  {W}_q   {\cal J}_{q,\omega}  \,, \\ 
		{\cal J}_{q,\omega}  & =  - \mbox{Im}\langle\!\langle(R^{2}-L^{2} ), (R^{2}-L^{2} )\rangle\!\rangle_{q,\omega} \,,  
		\end{aligned}
\end{equation}
Without interaction, $g_{q}=0$, and in the limit of bare bosons, $m^{-1}\to0$, the quantity ${\cal J}_{q,\omega} $ is given by 
    \begin{equation}
    {\cal J}_{q,\omega}  = - \frac{q(q^2+ (2\pi T/v)^2)}{48\pi^2}
        (\delta (qv - \omega) - \delta (qv + \omega) )    \end{equation}
see Eqs. (37), (38) in \cite{aristov2007}.  This expression produces corrections to $\sigma_{1}$ and should be ignored. 

\begin{figure}
\centering
\includegraphics[width=0.8\linewidth]{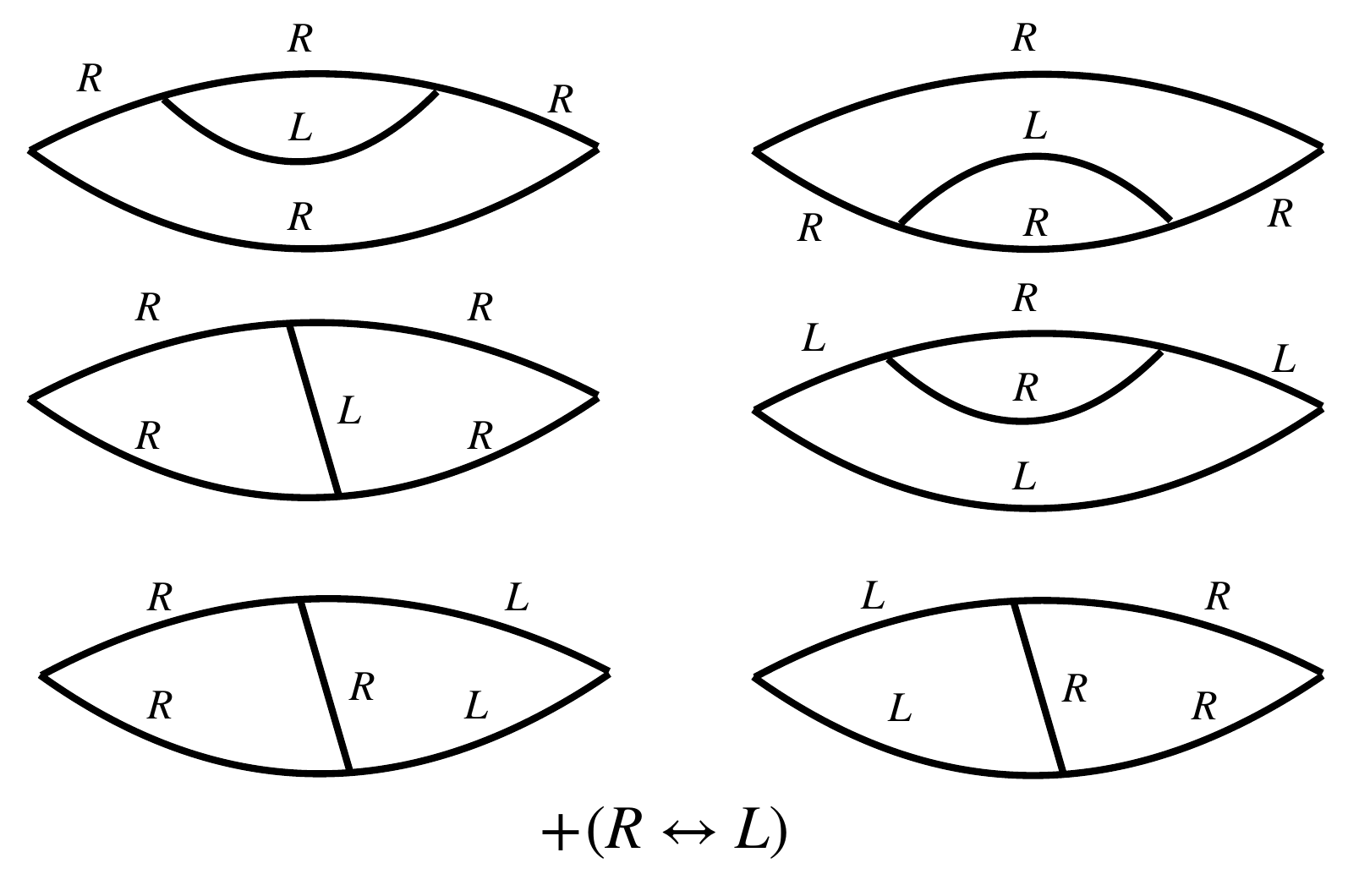}
\caption{ The diagrams in  second order of the vertex  $\Gamma^{(2)}$ for quantity ${\cal J}_{q,\omega} $  in  \eqref{eq:omega5} possessing the property ${\cal J}_{q \neq 0,\omega\to 0} \sim \omega $, and thus defining the asymptote $\sigma(\omega) \propto \omega^{-4}$. 
}
\label{fig:RRL}
\end{figure}

The first relevant corrections appear in presence of interaction,  $g_{q}\neq0$, which leads to appearance of $\Gamma^{(2)}_{0,0;0} \neq0$ in 
\eqref{eq:hamilBozed}.  We note that after the rotation \eqref{eq:Bogoliubov} the  quantity ${\cal J}_{q,\omega}$ preserves its form, because we may let approximately
$(R^{2}-L^{2} )= (\tilde R^{2}- \tilde L^{2} )$  in the considered case, $ d\gg a $.  In the second order of  $\Gamma^{(2)}$ we obtain several terms of the same order, depicted in Fig. \ref{fig:RRL}. Their spectral weight is characterized by the property ${\cal J}_{q \neq 0,\omega\to 0} \sim \omega $, which ensures the appearance of $\omega^{-4}$ contribution in $\sigma(\omega)$, Eq.\ \eqref{eq:omega5}. 

\begin{figure}
\centering
\includegraphics[width=0.4\linewidth]{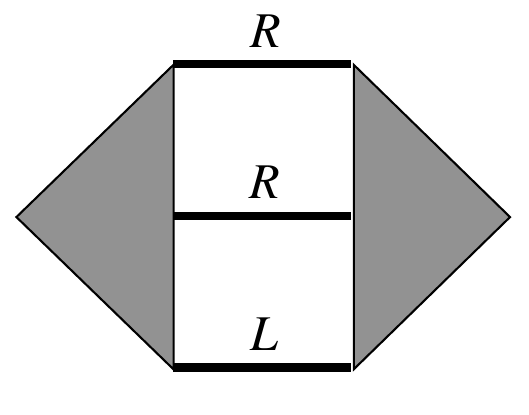}
\includegraphics[width=0.8\linewidth]{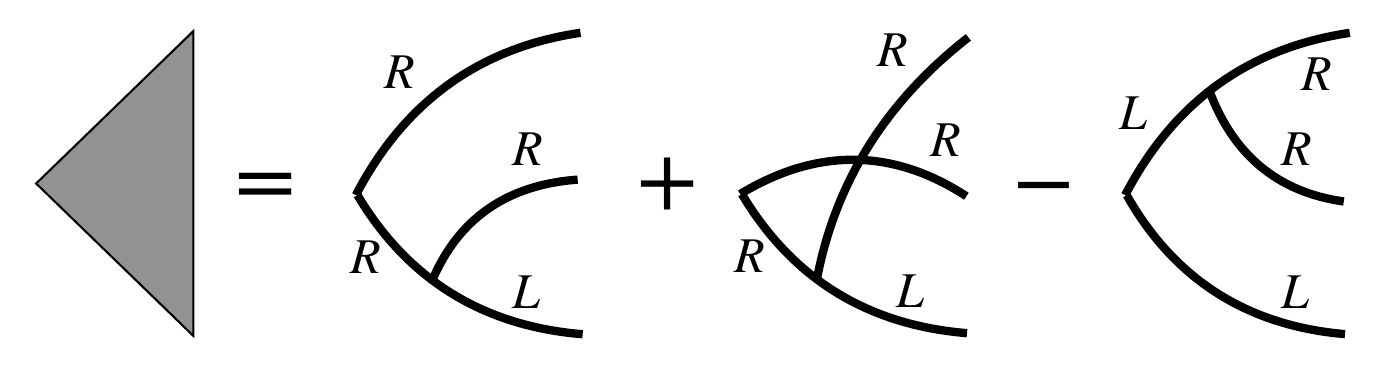}
\caption{ The diagrams in Fig.\  \ref{fig:RRL} are represented as two triangular vertices (gray shaded), connected by three quasiparticles in the cross-section, $S_{RRL}$. Those diagrams  in Fig.\  \ref{fig:RRL}, which are not shown and obtained by interchanging $R$ and $L$, correspond to cross-section  $S_{LLR}$, obtained similarly.  The analytical expressions for   $S_{RRL}$ and the total contribution are given by Eq. \eqref{Csection0}. 
}
\label{fig:RRL2}
\end{figure}

These terms partly compensate each other, and it is more convenient to sum them by the cross-section method, employed elsewhere in a similar problem.   \cite{aristov2007} This method represents the total contribution from the processes  in Fig.\ \ref{fig:RRL} as  two vertex parts, connected by three quasiparticles in the cross-section with an appropriate statistical weight, as shown in Fig.\  \ref{fig:RRL2}. 

 The general expression is (cf.\ Eq. (90) in \cite{aristov2007}): 
 \begin{equation}
 	\begin{aligned} 
{\cal J}_{q,\omega} & = -  \int do \;  |T_{jkl}|^2
 S_{jkl},
 \\
  do &= 4\pi^2 \delta\left[q-\sum q_i\right]
   \delta\left[\omega - \sum \Omega_i\right]
   \prod_{i=1}^3
  \left[\frac{dq_i d\Omega_i}{4\pi} \right]  ,\\
  S_{jkl} & = 
  \frac{g_{j}''(\Omega_1,q_1) g_{k}''(\Omega_2,q_2) g_{l}''(\Omega_3,q_3)}
 {\sinh\frac{\Omega_1}{2T}\;
 \sinh\frac{\Omega_2}{2T}\; \sinh\frac{\Omega_3}{2T}} \sinh\frac{\omega}{2T}\,,
% \\  & =   \frac{g_{j}''(x_1,q_1) g_{k}''(x_2,q_2) g_{l}''(x_3,q_3)}
 % {(2\sinh\frac{x_1}{2T})(2 \sinh\frac{x_2}{2T})(2 \sinh\frac{x_3}{2T})} 2\sinh\frac{\omega}{2T}\,,
 	\end{aligned} 
	\label{Csection0}
  \end{equation}
% 
%   \begin{eqnarray}
% {\cal J}_{q,\omega} & = & \pi  \int do \;  |\Gamma_{jkl}|^2 S_{jkl}\label{Csection0a}, \\
%  do &=&2\pi^2 \delta\left(\sum q_i\right)   \delta\left(\omega - \sum x_i\right)   \prod_{i=1}^3
%  \left[\frac{dq_i dx_i}{(2\pi)^2} \right]   \label{Csection0}  ,\\
%  S_{jkl} & =&   \frac{g_{j}''(x_1,q_1) g_{k}''(x_2,q_2) g_{l}''(x_3,q_3)} {T\sinh\frac{x_1}{2T}\;
% \sinh\frac{x_2}{2T}\; \sinh\frac{x_3}{2T}}, \nonumber \end{eqnarray}
%
here  the spectral weight for chiral density, $j =\tilde R, \tilde L$, is $g_{j}''(\Omega,q) = \frac{1}{2\pi i}(g_{j}(\Omega-i0,q) - g_{j}(\Omega+i0,q)) =
\pm (q/2\pi) K \delta(qv \mp \Omega)$, and the  
 Green's function,  $g_{j} (\Omega,q) = \frac{K}{2\pi} q/(qv \mp \Omega)$,
see Eq.\  \eqref{eq:rrLinearInteract}. 
% A symmetrization prefactor $1/2$ should be added to $S_{jkl}$ when values of two indices are coinciding.  

For the cross section with energies $\Omega_{1}$, $\Omega_{2}$, $\Omega_{3}$, referring to chiral propagators $R,R,L$, we have for the triangular vertex part 
shown in Fig. \ref{fig:RRL2} :
 \begin{equation}
 	\begin{aligned} 
 T_{RRL} &= \frac{4\pi^{2}}{m}\Gamma^{(2)}_{0,0;0} \left(
 g_{R}(\Omega_{2}+\Omega_{3},q_{2}+q_{3})  \right. \\ 
 &+  g_{R}(\Omega_{1}+\Omega_{3},q_{1}+q_{3})  \\
 &\left.  -  g_{L}(\Omega_{1}+\Omega_{2},q_{1}+q_{2}) \right ) \\
% & =\frac{4\pi^{2}}{m}  \frac{K}{2\pi v} \frac{q_{1}+q_{2}+q_{3}} { 2q_{3}} \Gamma^{(2)}_{0,0;0}  
& =   \frac{2\pi K}{ m v}  \Gamma^{(2)}_{0,0;0}  \frac{q_{1}+q_{2}+q_{3}} { 2q_{3}}  \,, 
 	\end{aligned}  
  \end{equation}
where we used the relations following from the spectral weight in the cross-section, $S_{RRL}$, and also $|q_{j}|\alt d^{-1}\ll a^{-1}$. 
In the limit of small $\omega$ we may let $\sinh({\omega}/{2T}) \simeq {\omega}/{2T}$ in \eqref{Csection0} and let $\omega = 0$ in the remainder there.  The diagrams with the cross-section $S_{LLR}$ double the contribution from those with $S_{RRL}$. 
Some calculation leads then to an expression  
 \begin{equation}
 	\begin{aligned} 
%{\cal J}_{q,\omega} & =  \frac{\omega}{\pi v^{3}} \left(\frac{K}{8\pi} \right)^{5}  \frac{q^{2}(q^2+ (4\pi T/v)^2)}{3T \sinh^{2}\frac{q v}{4T}} 	
{\cal J}_{q,\omega} & = - \frac{\omega K^{5} ( \Gamma^{(2)}_{0,0;0})^{2}}{1024 \,\pi^{2} m^{2} v^{3}}   \frac{q^{2}(q^2+ (4\pi T/v)^2)}{3T \sinh^{2}\frac{q v}{4T}} \,.
 	\end{aligned}  
  \end{equation}
Using now Eq.\ \eqref{eq:omega5}, we find $1/\omega^{4}$ asymptote of the conductivity in the leading  order of interaction and disorder. 
 
The coefficients in the expansion  \eqref{sigma-expa} read 
 
%   \begin{equation}  	\begin{aligned} 
%   x \gamma &= \frac{2   \left| \Gamma_{0,0;0}^{(2)} \right|^2 W_{0}}{v } \left[\frac{ T}{ \tilde{E}_F} \right]^{4} 
% \int \frac{dy \, y^4    }{ \sinh^2 y} \frac{  W_{4y T/v} }{W_{0} }  \,, \\
% x \gamma^{3} & = \frac{T^{2} K^{4}  \left| \Gamma_{0,0;0}^{(2)} \right|^2 W_{0}}{3v } \left[\frac{ T}{ \tilde{E}_F} \right]^{6} 
% \int \frac{dy \, y^6(y^{2} +\pi^{2})    }{ \sinh^2 y} \frac{  W_{4y T/v} }{W_{0} }  \,, 
% 	\end{aligned}    \end{equation}
  
  \begin{equation}
 	\begin{aligned} 
   A &=    \epsilon_{0} \left[\frac{ T}{ \tilde{E}_F} \right]^{4}  F_{1}(4 T/v)\,, \\ 
%    x \gamma &=    \epsilon_{0} \,( { T}/{ \tilde{E}_F} )^{4}  F_{1}(4 T/v)\,, \\ 
% \int \frac{dy \, y^4    }{ \sinh^2 y} \frac{  W_{4y T/v} }{W_{0} }  \,, \\
 B & =  \epsilon_{0}  K^{4} T^{2} \left[\frac{ T}{ \tilde{E}_F} \right]^{6} F_{2}(4 T/v)\,, \\
% x \gamma^{3} & = \tfrac{ 2  }{3  } \epsilon_{0}  K^{4} T^{2} ( { T}/{ \tilde{E}_F} )^{6} F_{2}(4 T/v)\,, \\
% \int \frac{dy \, y^6(y^{2} +\pi^{2})    }{ \sinh^2 y} \frac{  W_{4y T/v} }{W_{0} }  \,, \\
\epsilon_{0} & =  \left| \Gamma_{0,0;0}^{(2)} \right|^2 \frac{W_{0} }{v {\color{black}  K^{4} }} =  \frac{(K-1)^2 W_{0} }{16 \, v_{F} {\color{black}  K^{4} }} \,,  \\ 
F_{1}(q) & = \int \frac{dy \, y^4    }{ \sinh^2 y} \frac{  W_{q y} }{W_{0} }  \,, \\
F_{2}(q) & =\frac{ 2  }{3  }
\int \frac{dy \, y^6(y^{2} +\pi^{2})    }{ \sinh^2 y} \frac{  W_{q  y} }{W_{0} }  \,, \\
 	\end{aligned}  
	\label{xgamma1}
  \end{equation}
with renormalized Fermi energy $ \tilde{E}_F = m v^{2}/2$, $v=v_{F}/K$ and we used again $d \gg a$, so that at momenta $|q| \alt d^{-1}$ the value of $K_{q}$ is already saturated to $K=K_{0}$. The quantity $\epsilon_{0}$ is an energy scale, combining the strength of disorder and the electron interaction. 

\subsection{Redistribution of optical weight}

 The solution of the set of equation $A=\zeta\gamma$ and $B=\zeta \gamma^3$, where parameters $A$ and $B$ are given by (\ref{xgamma1}), is  
  \begin{equation}
 	\begin{aligned} 
\gamma  & = K^{2} \frac{ T^{2}}{ \tilde{E}_F} \sqrt{\frac{F_{2}(4 T/v)}{F_{1}(4 T/v)} } \,, \\ 
   \zeta  &=    \frac{  \epsilon_{0} T^{2}}{ \tilde{E}_F^{3} K^{2}}  \sqrt{\frac{F_{1}^{3}(4 T/v)}{F_{2}(4 T/v)} } \,.  
 	\end{aligned}  
	\label{xgamma2}
  \end{equation}

The behavior of $F_{1,2}(q)$ in opposite limits of its argument is  
%  \begin{equation} 	\begin{aligned} 
% F_{1}(q) & =  \pi^{4} /30 \,, \quad q \ll d^{-1}  \,, \\
% & \propto (q d )^{-3} \,,  \quad q \gg d^{-1}  \,, \\
% F_{2}(q) & =4\pi^{8} /105 \,,  \quad q \ll d^{-1}  \,, \\
% & \propto (q d )^{-5}  \,,\quad q \gg d^{-1}  \,, \\
% 	\end{aligned}    \end{equation}
    \begin{equation}
 	\begin{aligned} 
F_{1}(q) & =  \pi^{4} /30 \,, \quad F_{2}(q)   =4\pi^{8} /105 \,,  \quad q \ll d^{-1}  \,, \\
F_{1}(q) & \propto (q d )^{-3} \,,  \quad  F_{2}(q) \propto (q d )^{-5} \,,  \quad q  \gg d^{-1}  \,. 
 	\end{aligned}  
  \end{equation}
Therefore,  we have the following expressions for lower and higher temperatures: 
   \begin{equation}
 	\begin{aligned} 
\gamma  & = \pi^{2} K^{2} \frac{ T^{2}}{ \tilde{E}_F} \sqrt{8/7 } \,, \quad T \ll \omega_{d}    \,, \\ 
             & \sim K^{2} \frac{ T \omega_{d} }{ \tilde{E}_F}  \,, \quad  T \gg \omega_{d}    \,, \\ 
   \zeta  &=    \frac{ \pi^{2} \epsilon_{0} T^{2}}{ 60 \tilde{E}_F^{3} K^{2}}      \sqrt{7/2 } \,, \quad  T \ll \omega_{d}   \,. \\ 
    & \sim  \frac{   \epsilon_{0} \omega_{d}^{2}}{   \tilde{E}_F^{3} K^{2}}   
    \,, \quad  T \gg \omega_{d}    \,, \\ 
 	\end{aligned}  
	\label{xgamma3}
  \end{equation}

These expressions have several remarkable properties. Firstly, the width, $\gamma$, of the peak in $\sigma(\omega)$ at all temperatures does not contain the amplitude of the disorder potential, $W(q=0)$, but only its scale, $d$, and the curvature. Its dependence on the interaction, $g$, is  weak, as it enters through the overall prefactor of $K^{2}$. Further, one has $\gamma \sim (T/E_{F}) \min[ T , \omega_{d}] \ll \min[ T , \omega_{d}]$, i.e.\ $\gamma$ is the smallest energy scale in our problem.  It agrees with our assumptions in derivation of Eq.\ \eqref{sigma-expa}. 

Secondly, the part of the optical weight, $\zeta$, transferred from the $\delta$-function part of $\sigma(\omega)$, is explicitly proportional to the amplitude of disorder,   $W(0)$, the strength of  electron interaction  and the curvature. Interestingly, the optical weight, $\zeta$,  at higher temperatures, $T>\omega_{d}$, is saturated to a constant value, depending  on the disorder parameters, curvature and interaction strength. 

Thirdly, the value of $\sigma_{2}(\omega)$ at its  maximum, $\omega=0$, is 
   \begin{equation}
 	\begin{aligned} 
    \frac{  e^2 v_F}{\pi \hbar} \frac{\zeta}{\gamma}  & = \frac{  e^2 v_F}{\pi \hbar} 
     \frac{\epsilon_{0} F_{1}^{2} }{  {E}_F^{2} F_{2} }   \,,     \\  
     &\sim    \frac{  e^2  }{\pi \hbar}  \frac{ W_{0}   }{  {E}_F^{2}  } (K-1)^{2} {\cal O}(1)    \,, \quad  T \ll \omega_{d}   \,, \\ 
    & \sim     \frac{  e^2  }{\pi \hbar}  \frac{ W_{0}  }{  {E}_F^{2}  } \frac{\omega_{d}}{T}  (K-1)^{2} {\cal O}(1)
    \,, \quad  T \gg \omega_{d}    \,. \\ 
     \end{aligned}  
     \label{inelastic-max}
\end{equation} 
Comparing it with the elastic contribution,  $\sigma_{1} \simeq \frac{e^{2}  W_0}{ \pi \hbar E_F^{2}} $, Eq.\ \eqref{eq:sig1Inter}, we see that the inelastic part, $\sigma_{2}(\omega)$, is always smaller than the elastic one in the limit of small $\omega$.

\subsection{The zero frequency limit}

We saw above that in the leading order of smooth disorder potential the main part of the optical weight remains concentrated within the $\delta$-function peak at zero frequency.  One may ask a question, under which conditions this $\delta$-function peak may acquire a finite width and height. 

In this regard, let us discuss the problem of the conductivity of a one-dimensional degenerate electron gas with smooth disorder in a constant electric field. If the magnitude of the random potential is much smaller than the Fermi energy and its spatial scale is much larger than the Fermi wavelength, then the probability of backscattering of electrons on this potential is exponentially small, $\exp(-2k_Fd)$, where $k_F$ is the Fermi momentum. 

Being exponentially suppressed, these backscattering processes are nevertheless important in the limit $\omega \rightarrow 0$, when the d.c.\ conductivity becomes proportional to the length of the backscattering-related electron's path. In addition to such direct backward scattering with momentum transfer $2k_F$, there is an alternative process of momentum relaxation associated with the collision of two or more electrons with simultaneous forward scattering of one of them on the disorder. The momentum transferred to the electronic system in this process is of the order of $1/d \ll k_F$, which at $d \gg \lambda_F$ does not directly lead to a finite resistance.  

However, such collisions make possible the diffusion of  electrons in momentum space from the vicinity of point $k_F$ to the vicinity of point $-k_F$ through the bottom of the spectrum by elementary steps of order $1/d$, which leads to finite resistance.  \cite{levchenko2010transport} The momentum transfer to the diffusing electron occurs as it collides with the electrons in the band of the width of temperature near the Fermi level. Deep below the Fermi level, the fraction of unoccupied states at the energy $\epsilon$ is proportional to $\exp(-\epsilon/T)$, so the probability of such a diffusion transition is ultimately proportional to $\exp(-E_{F}/T)$. 
 
Comparing this probability with the above-mentioned probability of backward scattering, we conclude that the diffusion process is more effective under the condition $\exp(-E_F/T) \gg \exp(-2k_Fd)= \exp(-4E_F/\omega_{d})$, i.e. at temperatures $T\agt \omega_{d}$, and leads to conductivity proportional to $\exp(E_{F}/T)$. 
Given the the field frequency is smaller than the electron diffusion rate through the bottom,  this momentum relaxation mechanism should work also at $\omega \neq 0$. 
%It is clear that this momentum relaxation mechanism “works” also at $\omega \neq 0$, if the period of the field change is greater than the time of electron diffusion through the bottom. 
Thus, instead of a $\delta$-function at zero, the function $\sigma(\omega)$ has an exponentially narrow high peak with a maximum at zero.
{\color{black}  The optical sum rule dictates, that the width of this peak is inversely proportional to its height, i.e. is exponentially small. }

It should be stressed, that the described mechanism of broadening of $\delta$-function peak depends only on the electron interaction and happens even without disorder, in contrast to two contributions, $\sigma_{1,2}$, explicitly proportional to disorder strength. 

%%%%%%%%%%%%%%%%%%%%%
% $-------------------$
%%%%%%%%%%%%%%%%%%%%% 

\section{\label{sec:Conclusions} Conclusions}

We discussed the conductivity of  electron system in a sample with  random impurity potential, focusing on  the case of a large-scale potential, when backward scattering on it can be neglected. 
The conductivity at  finite frequencies is non-zero solely due to the curvature of fermionic dispersion. 
Using  bosonization techniques, we find conductivity for non-interacting electron gas  as well as for liquid with interaction of finite radius.  
Comparing our results with the analysis of classical gas without particle interaction, 
we establish the correspondence of two methods.   

% Within the framework of the bosonization method, we obtain a general formula for the conductivity of an electron system in a sample with a random impurity potential, then we apply the formula to the case of a large-scale potential $d\gg a$, when backward scattering can be neglected. Then we present several specific limiting cases of the formula. 

%The calculations are done for a non-interacting gas as well as for a liquid with a finite radius interaction.  We compare the results to the ones for a classical gas with no particle interaction, providing insights into the physical mechanism behind the conductivity and energy dissipation in this system.

%Throughout the paper we use the bozonization technique, which imposes the only important limitation: the frequency must significantly be high enough. Despite minor constraints, the bosonization technique allows us to avoid some common constraints. For instance, we do not assume that the interaction is either point-like or weak. 

We find that in the leading order of disorder potential strength, the conductivity is decomposed into three contributions, the $\delta$-function peak at zero frequency, and  two other contributions $\sigma_1(\omega)$ and $\sigma_2(\omega)$, appearing due to disorder potential, which we refer to as elastic and inelastic contribution correspondingly. The elastic contribution, $\sigma_1(\omega)$, does not depend on temperature and weakly depends on frequency and interaction;  it survives even in non-interacting case and is exponentially small at $\omega$ beyond the disorder-related scale $\omega_{d}$.  

 The inelastic contribution, $\sigma_2(\omega)$, occurs  in  simultaneous presence of disorder and interaction. It strongly depends  on temperature, frequency and interaction, and we analyze this dependence to some detail. 
It turns out that  $\sigma_{2}$ dominates over $\sigma_1$ only at high frequencies,  when $\sigma_{2} \sim |\omega|^{-5}$.  

We show that the overall weight of $\sigma_{1,2}(\omega)$ is proportional to disorder strength. Using the optical sum rule, we conclude that the weight of the $\delta$-function peak remains almost intact for weak disorder. 

%We show that    $\sigma_{2}$ is dominant over   $ \sigma_1 $ only at higher frequencies.  
%The overall conductivity consists of two contributions $\sigma_1(\omega)$ and $\sigma_2(\omega)$ that we refer to as elastic and inelastic contribution correspondingly. Conductivity $\sigma_2(\omega)$ decreases with frequency as a power of $1/\omega$ and $\sigma_1(\omega)$ decreases exponentially with $\omega$. Additionally, inelastic conductivity $\sigma_2(\omega)$ depends strongly on temperature, frequency and interaction, while $\sigma_1(\omega)$ does not depends  on temperature and weakly depends on frequency and interaction. 

%The inelastic contribution $\sigma_2$ is particularly sensitive to the nonlinearity of the spectrum, emphasizing the importance of nonlinearity in energy dissipation and conductivity. Indeed, the simple linear Luttinger liquid cannot dissipate energy even in the presence of the random potential.

%Remarkably, the inelastic contribution $\sigma_2$ demonstrates distinctive dependence on the frequency such as a $1/\omega^5$ dependence in (\ref{eq:wTvd}) and (\ref{eq:wvdT}); a $1/\omega^6$ dependence in (\ref{eq:Twvd}); and, within a specific interval, it even exhibits an $\omega^2$ dependence in (\ref{eq:vdwT}).

\section*{Acknowledgments}
We thank I.V. Gornyi and D.G. Polyakov for numerous helpful discussions. 
The work of D.N.A. was carried out with financial support from the Russian Science Foundation (grant No. 25-12-00212).

\appendix

{\color{black} 
\section{\label{AppOPE}  Bosonization basics }

%We use the bosonic primary chiral fields, including their zero mode part. 
% We write  $\bar \phi_{R}(x) = \int^{x} dx' \,R(x')$ , 
% $\bar \phi_{R}(x) =  k_{F} x +\phi_{R}(x)$ , $\bar \phi_{L}(x) = - k_{F} x +\phi_{L}(x)$. 
% The bosonization representation of fermion operators, $\psi_{R} = :\exp (i \bar \phi_{R}(x)):$ , 
% $\psi_{L} = :\exp (i \bar \phi_{L}(x)):$  is justified by the 
% short-range operator product expansions
% \begin{equation} \begin{aligned}
%  &  :\exp (i \bar \phi_{R}(x)):  \,  :\exp (-i \bar \phi_{R}(y)) :  \\ 
% & = \frac{}{}. :\exp (i  (\bar \phi_{R}(x)) - \bar \phi_{R}(y)) :  
% \end{aligned} \end{equation}
We remind the fermion-boson correspondence formulas known in the bosonization technique 
\cite{Giamarchi2003} 
\begin{equation}
\begin{aligned}
% \frac{1}{2m} \psi_{R}^{\dagger} (-i\nabla)^{2}\psi_{R}& 
% =  \frac{2\pi^{2}}{3m} R^{3}  \,, \quad
%  \frac{1}{2m} \psi_{L}^{\dagger} (-i\nabla)^{2}\psi_{L} 
% =  \frac{2\pi^{2}}{3m} L^{3}  \,, \\ 
  \psi_{R}^{\dagger} \frac{(-i\nabla)^{2}}{2m}\psi_{R}& 
=  \frac{2\pi^{2}}{3m} R^{3}  \,, \quad
  \psi_{L}^{\dagger} \frac{(-i\nabla)^{2}}{2m}\psi_{L} 
=  \frac{2\pi^{2}}{3m} L^{3}  \,, \\ 
 \psi_{R}^{\dagger} (-i\nabla)\psi_{R}& 
=  \pi   R^{2}  \,, \quad
  \psi_{L}^{\dagger} (-i\nabla)\psi_{L} 
= -\pi   L^{2}  \,, \\ 
 \psi_{R}^{\dagger}  \psi_{R}& 
=    R   \,, \quad
 \psi_{L}^{\dagger}  \psi_{L} 
=   L   \,, \\ 
\end{aligned}
\end{equation}
It shows that the fermionic gas Hamiltonian in presence of electromagnetic field, $\mathbf{A}(x)$, can be written as
\begin{equation}
\begin{aligned}
H_{0}&=\frac{1}{2m} \psi ^{\dagger}  (-i\nabla-e \mathbf{A}/c)^{2} \psi  \\
&=  \frac{2 \pi^{2}}{3  m} 
\left( \Big(R - \frac{eA_{\|}} {2\pi c} \Big)  ^{3}  
+ \Big(L + \frac{eA_{\|}} {2\pi c} \Big)  ^{3}  \right)
\,,  \\  
\end{aligned}
\end{equation}
where $A_{\|}$ is the component of $\mathbf{A}(x)$ along the wire. 
Note that any static $A_{\|}(x)$ may be absorbed into redefinition of $R$, $L$ by a shift, $R\to R +\tfrac{eA_{\|}} {2\pi c} $, and similarly for $L$,   which corresponds to gauge transform of  fermions, $\psi  \to \psi  \exp(i\tfrac{e} {c}\int ^{x} dx' A_{\|}(x')) $.
We do not consider the ring geometry  and finite magnetic flux through this ring, which would require the discussion of  persistent currents.
The smooth disorder potential $U(x)$ and the chemical potential $\mu$ can be added here,  by writing  
\begin{equation}
\begin{aligned}
 H_{0} + H_{U} & \to H_{0} + (U(x)- \mu) \psi ^{\dagger}   \psi 
 \\ &  =
H_{0} + (U(x)- \mu)  (R+L)   \,,
\end{aligned}
\end{equation}
Let us set $A_{\|}(x) \equiv 0$ and minimize  $H_{0} + H_{U}$ with respect to $R$, $L$. The found static solution, 
$\langle R\rangle =\langle L\rangle = [2m(\mu - U(x))]^{1/2} /(2\pi) \leftrightarrow k_{F}/(2\pi)$ corresponds to a non-oscillating  part of the fermionic density, around which the fluctuating part should be considered. 
We consider the electrically neutral setup, when the interaction of fermions happens only between the fluctuating parts of the density. 
In the limit of small $U(x)$ we write $\langle R\rangle=\langle L\rangle   \simeq k_{F}/(2\pi) - U(x)/(2\pi v_{F})$. Absorbing constant $k_{F}/(2\pi)$ into the redefinition of $R$, $L$, we return to  Eq. \eqref{eq:HamilDens} in the main text. 

Notice also that the disorder potential $U(x)$ and vector potential $A_{\|}(x)$ couple differently to chiral fermionic densities, so that variation of $H_{0}$ over $U(x)$ gives the density, $\rho= (R+L)$, whereas the variation over $A_{\|}(x)$ gives the current $j = -c\, \delta H_{0} /\delta A_{\|} \to\tfrac {\pi e }{m }(R^{2}-L^{2}) - \tfrac{e^{2}}{mc} A_{\|} (R+L)$, which after shifting  $R,L$ by $k_{F}/(2\pi)$ and neglecting $A_{\|}$ becomes  Eq.\ \eqref{eq:CurrentDiff}.  
 
It may be useful to provide here the derivation of the complex-valued conductivity in the clean non-interacting  case ($U(x)=0$, $g(x)=0$), starting from the usual Kubo formula instead of \eqref{eq:Kubo}, namely
 \[
	\sigma(\omega)= \frac{i  } {\hbar\, \omega }( \langle\!\langle   j ,\,   j  \rangle\!\rangle _{ \omega} 
	- \langle\!\langle   j ,\,   j  \rangle\!\rangle _{ \omega=0}) \,,
\]
taken at $q=0$. The second term in parenthesis arises from the diamagnetic term, neglected above, whose averaging gives,  $- \tfrac{e^{2}}{mc} A_{\|} \langle R+L \rangle = - \tfrac{e^{2}}{\pi c} A_{\|} v_{F}$. This quantity  coincides with $- \tfrac{1}{ c} A_{\|} 
\langle\!\langle  {  j ,\,   j} \rangle\!\rangle _{\omega=0}$ in the limit $q=0$, which reflects the Ward identity corresponding to charge conservation.   The retarded function $\langle\!\langle  {  j ,\,   j} \rangle\!\rangle _{q,\omega}$ is given by $\langle\!\langle  {  \rho ,\,   \rho} \rangle\!\rangle _{q,\omega}$, Eq.\ \eqref{eq:rrNoInt}, up to a factor, 
\[
\langle\!\langle  {  j ,\,   j} \rangle\!\rangle _{q,\omega}= \frac{1}{2\pi}\frac{v_{F}^{3} q^{2}}{(\omega+i0)^{2}-v_F^{2}q^{2} } \,, 
\]
so that the Kubo formula reads 
\[
\begin{aligned}
	\sigma(\omega) & =\lim_{q\to0} \frac{i v_{F} } {2\pi   } \left(  \frac{1} {\omega+i0 - v_F q} + \frac{1} {\omega+i0 + v_F q} \right) 
	\\ & 
	=  \frac{i v_{F} } { \pi   }  \frac{1} {\omega+i0  }   
	\,,
\end{aligned}
\]
and the real part of conductivity is $\sigma(\omega) = v_{F} \delta(\omega)$. 

Let us also briefly discuss the notion of  ``point-like interaction'', mentioned after Eq.\ \eqref{eq:GeneralCorrel}. Starting from the fermionic representation of interaction, $  \psi^{\dagger} (x) \psi(x)   g(x-y) \psi^{\dagger} (y) \psi(y) $, we pass to Fourier transform,  
$\psi(x) =  e^{i k_{F}  x} \psi_{R}(x) +e^{-ik_{F} x} \psi_{L}(x) = \sum_{p} ( e^{i (k_{F}+p) x} \psi_{R,p} +e^{i (-k_{F}+p) x} \psi_{L,p})$, 
and find two combinations,  one of them $ (\psi_{R}^{\dagger}(x)\psi_{R}(x)+ \psi_{L}^{\dagger}(x)\psi_{L}(x))   g(x-y) 
 (\psi_{R}^{\dagger}(y)\psi_{R}(y)+ \psi_{L}^{\dagger}(y)\psi_{L}(y)) $ is already of the form Eq.\ \eqref{eq:HamilDens}. Another combination,  
 $ \psi_{R}^{\dagger}(x)\psi_{L}(x)  \psi_{L}^{\dagger}(y)\psi_{R}(y)e^{2i k_{F} (x-y)}   g(x-y)  + H.c. $,   adds $  g_{q\simeq 2k_{F}}$ to the mixed term of the interaction, $g_{2}RL $.  % component of the interaction between the smooth densities,  $R$ and $L$.  
Since we assume that the range of the interaction is large, $a\gg \lambda _{F}$, we expect that the latter combination, $  g_{q\simeq 2k_{F}}$, is (exponentially) small compared to $  g_{q \simeq 0}$ and  may be neglected. In this sense, the ``point-like interaction''  means  that $a$ is small compared to all other scales, except for $ \lambda _{F}$.  
}

\section{\label{sec:NonIntClassi} One-particle contribution obtained with the classical Liouville's equation }	
For brevity, we omit index 'classical' in this appendix. We consider the linear response of the classical system and may replace $f_{R}$ in the right-hand side part of (\ref{eq:KinurClass}) with the Fermi distribution function of the right electrons, $f^{(0)}_{R}$, and $f_{R}$ in the lefthand side  now means deviation from it, $  f^{(1)}_{R} = f _{R} - f^{(0)}_{R}$.  

The Green's function for Eq.\ (\ref{eq:KinurClass}) then reads
\begin{equation}
\begin{aligned}
\label{eq:KinurClass2}
G(x, E, t \mid & x', E', t') = \frac{ \theta(t-t') \delta(E-E')}{\sqrt{2\left(E' - U(x')\right) / m}} 
\\
& \times  \delta\left(t - t' - \int_{x'}^{x} \frac{ {d}x_1}{\sqrt{2\left(E' - U(x_1)\right) / m}}\right) \,.
\end{aligned}
\end{equation}
In the zero-temperature limit $f^{(0)}_{R}(E)=\theta(\mu-E)$, so that 	
\begin{equation}
\label{eq:KinurClass3}
\frac{\partial f_{F, R}}{\partial E} = -\delta(\mu - E) \,.
\end{equation}

Expressions (\ref{eq:KinurClass}), (\ref{eq:KinurClass2}) and (\ref{eq:KinurClass3}) give the correction to $f^{(0)}_{R}$   from the external field ${\cal E}_{0}(t)$ :
$$
\begin{aligned}
f^{(1)}_{R} (x, t) &= -e \delta(E - \mu) \int_{-\infty}^{x}  {d}x' \,
\\ &\times
{\cal E}_{0} \left [ t-  \int_{x'}^{x} \frac{ {d}x_1}{\sqrt{2\left(E - U(x_1)\right) / m}} \right ]  \,.
\end{aligned}
$$ 
The field-induced flux of the right particles at $x$ is obtained by (i) multiplying $g_{R}(x, t)$ by $-e$ and velocity  $V=\sqrt{2(E-U(x)) / m}$ and (ii) integrating it over the energy with the density of states 
$\nu(E, x)=\frac{1}{2 \pi \hbar \sqrt{2(E-U(x)) / m}}$.
%  $\nu(E, x)=\frac{\theta(E-U(x))}{2 \pi \hbar \sqrt{2(E-U(x)) / m}}$.

\begin{widetext}
$$
j_{R}(x, t)=\frac{e^{2}}{2 \pi \hbar} \int_{-\infty}^{x} d x^{\prime}\,  {\cal E}_{0}\left[t-  \int_{x'}^{x} \frac{ {d}x_1}{\sqrt{2\left(\mu - U(x_1)\right) / m}} \right ]     \,. 
$$	
After averaging  (denoted by angular brackets in this section)  this expression over realizations of the random potential the coordinate dependence vanishes, $\left\langle j_{R}(x, t) \right\rangle = \left\langle j_{R}(0, t) \right\rangle $,  and the flux of the right particles  becomes
$$
\langle j_{R}(0, t)  \rangle = \frac{e^{2}}{2 \pi \hbar} \int_{-\infty}^{0}  {d}x  \left\langle {\cal E}_{0}\left [ t-  \int_{x}^{0} \frac{ {d}x_1}{\sqrt{2\left(\mu - U(x_1)\right) / m}} \right]   \right\rangle    \,.  $$
Repeating similar steps for left-moving electrons, we find 
$$
\langle j_{L}(0, t) \rangle = \frac{e^{2}}{2 \pi \hbar} \int_{0}^{\infty} {d}x   \left\langle {\cal E}_{0}\left[t-  \int_{0}^{x} \frac{ {d}x_1}{\sqrt{2\left(\mu - U(x_1)\right) / m}} \right]   \right\rangle    \,,  $$
with the total current $j(t) =  \langle j_{R}(0, t)  \rangle + \langle j_{L}(0, t) \rangle$.

The conductivity is obtained from the relation $j(t) =    \int dt'\,\sigma(t-t') {\cal E}_{0}(t')$, which gives 
\[ 
\sigma(t) = 
\frac{e^{2}}{2 \pi \hbar} \int_{-\infty}^{\infty}  {d}x   \left\langle \delta\left(t-   \int_{0}^{x} \frac{ \mbox{sign}\,x\, {d}x'}{\sqrt{2\left(\mu - U(x')\right) / m}} \right)   \right\rangle    \,.
\]
and in $\omega$-representation 
\begin{equation}
\label{eq:KinurClassSigmaGeneral}
\sigma (\omega)=\frac{e^{2}}{2\pi \hbar} \int_{-\infty}^{\infty}  {d}x  \left\langle \exp  i \omega   \int_{0}^{x} \frac{ \mbox{sign}\,x\, {d}x'}{\sqrt{2\left(\mu - U(x')\right) / m}}  \right\rangle \,.
\end{equation}

% \begin{equation} \label{eq:KinurClassSigmaGeneral} \sigma (\omega)=\frac{e^{2}}{\pi \hbar} \int_{-\infty}^{0} d x \left\langle \exp \left[-(i \omega - \gamma)    \int_{x}^{0} \frac{ {d}x_1}{\sqrt{2\left(\mu - U(x_1)\right) / m}}  \right] \right\rangle \,. \end{equation}

Integration over the  frequency gives the following sum rule
\begin{equation}
\begin{aligned}
\label{eq:KinurClassSigmaGeneralInteg}
\int_{-\infty}^{+\infty} \sigma_{\text{}}(\omega) d \omega & = 2\pi  \sigma(t=0) = \frac{e^{2}}{\hbar}\langle\sqrt{2[\mu-U(0)] / m}\rangle 
\\  & =
\frac{\left.e^{2} \pi\,\langle n\right\rangle}{2 m}\approx \frac{e^{2} v_{F}}{\hbar}\left(1-\left\langle U^{2}(0)\right\rangle / 8 \mu^{2}\right) 
\,,
\end{aligned}
\end{equation}
where $\langle n\rangle$ is the average concentration. % Here we take into account that the delta function $\delta(x)$ which appears during the calculation gives $1/2$ because of the integration limits. 
The last approximation is valid when $U(x)\ll \mu$. 

It is impossible to perform explicit averaging in (\ref{eq:KinurClassSigmaGeneral}). Instead we assume $U(x)\ll \mu$, keeping only the first and second order of $U / \mu$ in the series expansion. The result is	

\begin{equation}
\begin{aligned}
\label{eq:KinurClassSigmaGeneralAver}
\sigma(\omega)  & \simeq \frac{e^2}{2\pi \hbar} \int_{0}^{\infty}  {d}x 
\left\langle \exp   \left( i \omega  -\gamma \right) \int_0^x \frac{ {d}x'}{v_F} 
\left[ 1 + \frac{U(x')}{2 \mu} + \frac{3}{8} \left( \frac{U(x')}{\mu} \right)^2 \right] 
\right\rangle \\ 
&   +  \frac{e^2}{2\pi \hbar} \int_{-\infty}^0  {d}x 
\left\langle \exp   \left( -i \omega  +\gamma \right) \int_0^x \frac{ {d}x'}{v_F} 
\left[ 1 + \frac{U(x')}{2 \mu} + \frac{3}{8} \left( \frac{U(x')}{\mu} \right)^2 \right] 
\right\rangle \,, 
\end{aligned}
\end{equation}
with $\gamma \to +0$ to ensure convergence at large $|x|$.

%\begin{equation}  \label{eq:KinurClassSigmaGeneralAver} \sigma(\omega) \approx \frac{e^2}{\pi \hbar} \int_{-\infty}^0  {d}x \left\langle \exp \left\{ -\left( i \omega + \gamma \right) \int_x^0 \frac{\mathrm{d}x_1}{v_F} \left[ 1 + \frac{U(x_1)}{2 \mu} + \frac{3}{8} \left( \frac{U(x_1)}{\mu} \right)^2 \right] \right\} \right\rangle \end{equation}

Expanding the exponent   and averaging,  one obtains, e.g.,  for the first term in Eq.\ \eqref{eq:KinurClassSigmaGeneralAver} % (left-mover contribution)  
\begin{equation}
\begin{aligned}
\label{eq:KinurClassSigmaGeneralAverGauss}
\sigma(\omega) = & \frac{e^2}{\pi \hbar} \int_{0}^{\infty}  {d}x\, e^{ i \omega {x}/{v_F} } \left( 1  + i \omega \int_{0}^{x} {d}x\, \frac{3 W(x=0)}{8 v_F \mu^2}  - \omega^2 \iint_{0}^{x} {d}x_1  {d}x_2  \frac{W(x_1 - x_2)}{8 v_F^2 \mu^2}   \right) \,. 
% +(x_1\leftrightarrow x_2)\\=  
% & \frac{e^2}{\pi \hbar} \int_{0}^{\infty}  {d}x\, e^{ i \omega {x}/{v_F} } \left( 1  + i \omega \int_{0}^{x} {d}x\, \frac{3 W(x=0)}{8 v_F \mu^2}  - \omega^2 \iint_{0}^{x} {d}x_1  {d}x_2  \frac{W(x_1 - x_2)}{8 v_F^2 \mu^2}   \right),
\end{aligned}
\end{equation}
Here we used    $W(x-x')=\left\langle U(x) U(x') \right\rangle $ and the feature that $W(x)=W(-x)$ for an arbitrary $U(x)$.
After some algebra we can reduce it to the form  
% {\color{black}  \sout{
% \begin{equation} 
% \sigma(\omega)  \simeq \frac{e^2 v_F}{  \hbar} \delta(\omega)\left(1+ \frac{3 W(0)}{8 \mu^2} \right)^{-1}
% +  \frac{e^2 }{ \hbar}  \int_{-\infty}^{\infty}  {d}x\, e^{ i \omega {x}/{v_F} }\frac{ W(x)}{4  \mu^2}  \,  
% \end{equation}
% }
\begin{equation}
\begin{aligned}
\sigma(\omega) &= \frac{e^2 v_F}{\pi \hbar} \left\{ \frac{i}{\omega + i \gamma} \Big(1- \frac{3 W(x=0)}{8 \mu^2} \Big)
%\frac{i}{\omega + i \gamma} 
+ \frac{1}{4 v_F \mu^2} \int_{0}^{\infty} \exp \left(\frac{i \omega x}{v_F}\right) W(x) \,  {d}x \right\} \,, 
\\
Re \left[ \sigma(\omega) \right]  &=  \frac{e^2 v_F}{ \hbar} \left( 1 - \frac{3 W(x=0)}{8 \mu^2} \right) \delta(\omega) + \frac{e^2}{8 \pi \hbar \mu^2} W_{q=\omega/v_F} \,, 
\end{aligned}
\end{equation}
where $W_q=\int_{-\infty}^{\infty} \exp \left(i q x\right) W(x) dx$ is real-valued.
% , note that $W_q=W_q^*$ and $Re[W_q]=W_q$
Thus we obtain Eq. \eqref{eq:ClassResult}, and 
the above sum rule \eqref{eq:KinurClassSigmaGeneralInteg} is satisfied since
$$\int_{-\infty}^{+\infty} \sigma_{\text{}}(\omega) d \omega \approx \frac{e^{2} v_{F}}{\hbar}\left(1-W(0) / 8 \mu^{2}\right)\,. $$

% {\color{black} 
% After integration over x it reads
% $$
% \sigma(\omega) = \frac{e^2 v_F}{\pi \hbar} \left\{ \frac{-i}{\omega - i \gamma} + \frac{3 W(0)}{8 \mu^2} \frac{i}{\omega - i \gamma} + \frac{1}{4 v_F \mu^2} \int_{0}^{\infty} \exp \left(\frac{i \omega x}{v_F}\right) W(x) \,  {d}x \right\}.
% $$
% } 

%The real part of this expression is (\ref{eq:ClassResult}).

\end{widetext}

\bibliography{DDA_23_biblio}
\end{document}